\documentclass [12pt]{article}
\usepackage [cp1251]{inputenc}
\usepackage [english]{babel}
\usepackage {graphicx}
\usepackage {amssymb}
\usepackage {amsmath}
\usepackage {mathrsfs}
\usepackage {longtable}
\title{Gas-filled pore in bounded particle.}

\author{ $^{1,2}$\textbf{V.V. Yanovsky},$^{1}$\textbf{M. I. Kopp },$^{1}$\textbf{M. A. Ratner}}

\begin{document}

\maketitle

$^{1}$ \textit{Institute for single crystals, National Academy of
Science of Ukraine, Nauki Ave 60, 61001 Kharkiv, Ukraine}

$^{2}$\textit{Kharkiv National Kharazin University, Svobody sqr. 4, 61000, Kharkiv, Ukraine}

\abstract{The diffusive evolution has been studied of gas-filled pore has in a bounded particle in gas media. The nonlinear equation set, describing the behaviour of gas-filled pore on bounded particle is obtained. Asymptotic modes are considered for evolution of small and large pores. Analytical solutions are obtained in asymptotic modes. The comparison is conducted of these solutions with results of numerical solution of complete equation set. The characteristic regularities of  gas-filled pore behavior are found at arbitrary  pore position relative to matrix particle center.}

\section{Introduction}

One of the most important problems of contemporary material science is investigation of onset and development of gas porosity in materials. The creation of materials with improved radiation hardness is important for development of atomic energetics development as well as for other sectors of industry. Along with vacancy pores, gas-filled pores were discovered forming due to irradiating metals by quick neutron or charged particle fluxes in accelerators. For the first time, theoretical investigation of these problems was performed in the works \cite{1s}-\cite{7s}. In the same works, the growth of pores filled with noble gases was considered as applied to material swelling, that is, to a large degree, connected with pore coalescence. The physical cause of material swelling as a consequence of gas porosity consists in absorbing of thermal vacancies at redistribution of pores during the coalescence. Pore behaviour becomes even more complicated if it is filled chemically active gas (or gases)  that at coalescence temperatures can interact matrix material or other gases, forming inside the pore one or several gaseous compounds). Such situation can take place, for example, under irradiation. At that, fragments in the form of chemically active gas molecules are formed in the material. The process of gas-filled bubble formation can, probably, occur in many materials, since practically all real materials contain interstitial impurities in the form of oxide, carbide, nitride, and other phases \cite{4s}.

The phenomenon of gas porosity plays an important role in the investigation of the properties of metals and alloys under  irradiation by quick neutron or charged particle fluxes and at annealing of irradiated materials. Especially large threat is presented by the growth of pores that are filled with rare gases since it can restrict essentially the durability of materials in atomic reactors.

Because of this, the theory of pore ensemble evolution, its kinetics was described in the works \cite{1s}-\cite{7ss} in unbounded materials. In these works, the general theory is developed of diffusive decay of oversaturated solid alloys on the stage of coalescence, that is of the growth of larger nanodefects on the account of smaller ones. Fundamental equations for the description of vacancy and gaseous porosity were obtained in the works  \cite{4s},\cite{8s}-\cite{10ss}. It was shown in \cite{9s} that complicated system of nonlinear differential equations, that describes gas porosity, can be solved analytically in two limiting cases of small and large pores. Such conventional division of pores by sizes is connected with division of coalescence time into two stages. Thus, small pore size is tuned to the amount of gas inside the pore, and the pore starts to grow. For large pores, its filling by the gas is tuned to pore sizes, that changes insignificantly. Theory of gas porosity was generalized in the work \cite{10s} for the case of multicomponent systems.

Thus, at the present time, theory of vacancy and gas porosity is developed well for unbounded matrices \cite{1s}-\cite{10s}. However, the development of nanotechnologies requires new theoretical researches of defect structures in bounded particles of nano- and meso- scales. The most widespread three dimensional defects in such meso- and nanoparticles are vacancy pores, gas-filled pores as well as new phase inclusions. The regularities of diffusion growth, healing and motion of such defects in nanoparticles is an important problem. Such defect structure plays an important role for the possibility of further compactification of nanoparticles and creating new materials \cite{11s}. Establishing regularities of defect structure evolution will enable one to control it as well as to change properties of corresponding meso- and nanoparticles.

The creation of the theory of the diffusive evolution of pores in bounded medias, for example, in spherical nanoparticles, is a rather complicated task. 

In bounded particles of the matrix, the influence of close boundary complicates strongly pore behaviour.  Closeness of boundary leads to principally different pore behaviour as compared to that in unbounded materials. It is worse to note, that pore formation in spherical nanoshells was discovered relatively recently \cite{8s}. In the review \cite{13s} the results are presented of theoretical and numerical investigations related to formation and disappearing of pores in spherical and cylindrical nanoparticles. Great attention in \cite{13s} is paid to the problem of hole nanoshell stability, i.e. to the case when in the nanoparticle center large vacy pores are situated.  

Analytical theory of diffusive interaction of the nanoshell and the pore situated at arbitrary distance from particle center was considered in the work  \cite{14s}. With the supposition of quasiequilibrium of diffusive fluxes, the equations have been obtained analytically for the change of the radii of pore and spherical granule as well as of center-to-center distance between the pore and the granule. In \cite{14s}, the absence of critical pore size has been demonstrated unlike the case of inorganic matrix. In general case, pore in such particles dissolves diffusively, while diminishing in size and shifting towards granule center.

This problem is close to that of diffusion interaction of pores in unbounded matrix \cite{11ss}. Indeed, the role of second object the pore interacts with in bounded particles is played by matrix particle boundary. Interaction with boundaries leads to principally different pore behaviour as compared to that in unbounded materials. The formation of pores in spherical nanopaparticles was discovered experimentally in the work \cite{12s}. In the review \cite{13s} the results are presented of theoretical and numerical investigations related to formation and disappearing of pores in spherical and cylindrical nanoparticles. In \cite{13s}, great attention is paid to the problem of hole nanoshell stability, i.e. to the case when in the nanoparticle center large vacy pores are situated.  Analytical theory of diffusive interaction of the nanoshell and the pore situated at arbitrary distance from particle center was considered in the work  \cite{14s}. With the supposition of quasiequilibrium of diffusive fluxes, the equations have been obtained analytically for the change of the radii of pore and spherical granule as well as of center-to-center distance between the pore and the granule. In \cite{14s}, the absence of critical pore size has been demonstrated unlike the case of inorganic matrix. In general case, pore in such particles dissolves diffusively, while diminishing in size and shifting towards granule center.In the work \cite{15s}, the evolution of gas-filled pore was considered in a simple case of zero diffusion coefficient of the gas in the matrix. It has been shown that the behaviour of gas-filled pore is qualitatively different from that of vacancy pore in spherical matrix. Thus, unlike vacancy pore, there exist a stationary size of a gas-filled pore, that is determined by the gas density and granule temperature.  Asymptotic regimes as well as main regularities of the gas-filled pore behaviour has been established.

In the work \cite{15s}, the evolution of gas-filled pore inside spherical nanoparticle was considered in a simple case of zero diffusion coefficient of the gas in the matrix. It has been shown that the behaviour of gas-filled pore is qualitatively different from that of vacancy pore in spherical matrix. Thus, unlike vacancy pore, there exist a stationary size of a gas-filled pore, that is determined by the gas density and granule temperature.  In the work \cite{15s}, it was established that, in general case, gas-filled pore is shifts insignificantly towards granule center if its size decreases down to stationary value while the pore is shifted away from granule center if its size increases up to stationary value. 

In the present work, evolution of  gas-filled pore is considered inside spherical granule placed into gas media with account of diffusion of single atom gas and vacancies in the matrix. The main attention is paid to the case when initial gas pressure inside a pore is smaller then outer gas pressure $P_0>P_g(0)$. The nonlinear equation set is obtained, that determines evolution of position and size of the pore as well as gas density inside it and granule size. It is shown that in general case, gas pressure in a pore is increasing up to the value of outer gas pressure. Further evolution of the pore occurs at stationary gas pressure in a pore and consists in the decrease of its radius. The number of gas atoms in the pore at the initial stage is increasing until gas pressure in a pore rich the outer gas pressure. After that the number of gas atoms inside a pore decreases while pore is shifted towards granule center. Asymptotic regimes of small and large pores are considered that give simpler sets of nonlinear equations. Analytical solutions of these equations are obtained. The good agreement is shown of numerical solution of complete equation set describing  pore behavior with analytical solution under corresponding conditions. Life time of a  gas-filled pore is found analytically in all asymptotic modes, that agree well with those obtained at numerical solution of the complete equation set.

\begin{figure}
  \centering
	\includegraphics[width=15 cm, height=7 cm]{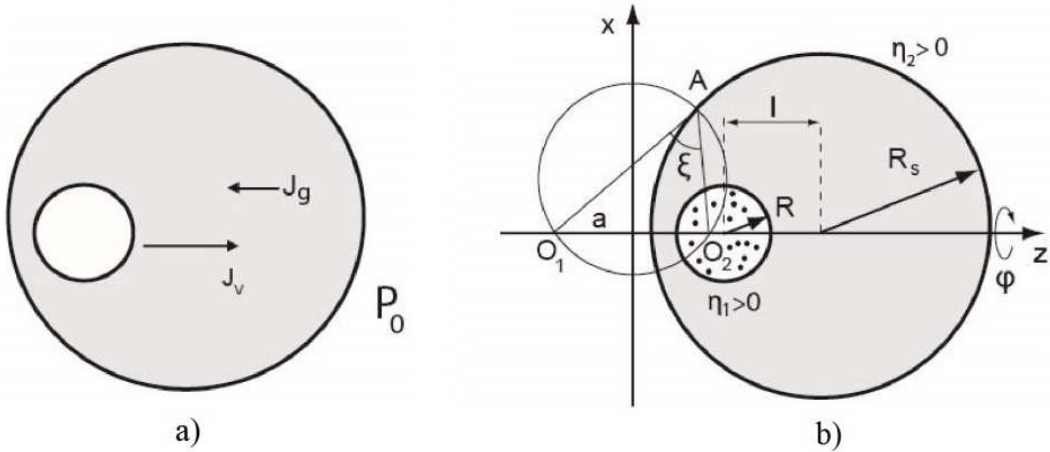}  \\
  \caption{a) The filling in of vacancy pore by gas is shown schematically. Gas pressure at granule boundary is constant $P_0$; b) Gas-filled pore inside spherical granule in bispherical coordinate system. Pore and granule surfaces in these coordinate system are coordinate planes  $\eta=\textrm{const}.$}  \label{fg1}
\end{figure}
The obtained results establish the character of evolution of gas-filled pore dependently on the basic characteristics of the matrix and the gas. This allows to predict change of  pore behavior dependently on the properties of the matrix and the gas that is important for detecting deviations from diffusion behavior at numerical modelling of nanoparticles with defect structures. It is just the comparison with diffusion behavior described in the present work that enables one to distinguish anomalous  behavior changes from regular ones.

\section{Statement of the problem and derivation of basic equations}

Let us consider the spherical granule of the radius $R_s$ containing the gas-filled pore of the radius $R < R_{s}$. Granule and pore centers are separated from each other by the distance $l$.  (see Fig. \ref{fg1}$\textrm{a}$).

There exist various mechanisms of gas atoms diffusion in the crystal lattice of a bounded matrix. The examples of such mechanisms are hopping of matrix atoms, as well as of complexes formed by dopant atoms with vacancies or by other ways. This influences the value of diffusion coefficient only. Thus, the obtained below results are of universal nature. Let us suppose, that granule is surrounded by single component chemically inactive gas under the pressure $P_0$. The gas can diffuse through granule material and fill in the pore. The complete description of such evolution assumes the knowledge of pore and granule size change with time as well as of time change of their center-to-center distances. In order to obtain the equations describing such evolution, the boundary conditions are required, that are determined by equilibrium vacancy concentrations near the pore and granule surfaces. The equilibrium vacancy concentration near a spherical pore surface is determined, with the account of gas pressure, by the relation see e.g. \cite{7s},\cite{16s} )
\begin{equation}\label{eq1}
    c_{R}^v=c_{V}\exp\left(\frac{2\gamma\omega}{kTR}-\frac{P_g\omega}{kT}\right),
\end{equation}
where $c_V$ is equilibrium vacancy concentration near the plane surface, $\gamma$ is surface energy, $T$ is granule temperature, $\omega$ is the volume per lattice site, $P_g$ is gas pressure inside the pore. For the sake of simplicity let us use state equation of ideal gas:
\[P\cdot \frac{4\pi}{3}\cdot R^3=N_g kT,\]
here $N_g$ is the  quantity og gas atoms inside the pore.
The deviation from an ideal case can be easily taken into account with the use of corresponding state equation.
In the same way, equilibrium concentration of vacancies near the spherical granule free surface is determined. 
\begin{equation}\label{eq2}
   c_{R_{s}}^v=c_{V}\exp\left(-\frac{2\gamma\omega}{kTR_s}-\frac{P_0\omega}{kT}\right),
\end{equation}
These concentration values will determine vacancy fluxes. Vacancy fluxes determine pore size, while gas fluxes determine gas pressure inside the pore and influence boundary conditions.
Stationary fluxes of vacancies inside granule are quickly established if the condition is met $\tau_v \ll R_s^2/D$, where $\tau_v$ is characteristic time of establishing stationary vacancy fluxes, $D$ is vacancy diffusion coefficient. It is natural to suppose that near granule boundary local thermodynamic equilibrium with the gas is established. Then relations (\ref{eq1})  and (\ref{eq2}) are complemented with boundary conditions for equilibrium gas concentration in the matrix near the surface of spherical pore as well as near free surface of spherical granule \cite{7s}:
\begin{equation}\label{eq3}
c_{R}^g=P_g\cdot \kappa_H, \quad c_{R_{s}}^g=P_0\cdot \kappa_H.\end{equation}
Here the coefficient $\kappa_H$ is used:
\[ \kappa_H=\frac{W_0}{kT}\cdot \delta, \]
where $W_0=(2\pi\hbar^2/mkT)^{\frac{3}{2}}\cdot \mathcal Z$, $\hbar$ is Plank constant, $m$ is the mass of atom or molecule,  $\mathcal Z$ is statistical sum over internal energy levels of molecule (for gas atom $\mathcal Z=1$). The value of $\kappa_H$ depends on whether gas atom is substitutional or interstitial one \cite{7s}. Here we will neglect the field of elastic strains, considering them small, therefore $\delta=\exp(-\psi/kT)$, where  $\psi$ is energy change of the system if one atom is added to the lattice.
Stationary fluxes of gas atoms in the granule also establish quickly enough if the condition is satisfied $\tau_g \ll R_s^2/D_g$, where $\tau_g$ is characteristic time of establishing stationary fluxes of gas atoms, $D_g$ is the coefficient of gas diffusion. Thus, in order to obtain equations for the change of the number of gas atoms in the pore we will use quasistationary fluxes.

The geometry of pore and granule boundaries dictates the use of bispherical coordinate system  \cite{17s}, as the most convenient one. In bispherical coordinate system  (see Fig.\ref{fg1}$\textrm{b}$) each point  $A$ of the space is matched to three numbers $(\eta,\xi,\varphi)$, where $\eta=\ln(\frac{|AO_1|}{|AO_2|})$, $\xi=\angle O_1AO_2$, $\varphi$ is polar angle.
Let us cite relations connecting bi-spherical coordinates with Cartesian ones: 
\begin{equation}\label{eq4}
  x=\frac{a\cdot \sin\xi\cdot\cos\varphi}{\cosh\eta-\cos\xi}, \quad
  y=\frac{a\cdot \sin\xi\cdot\sin\varphi}{\cosh\eta-\cos\xi}, \quad
  z=\frac{a\cdot \sinh\eta}{\cosh\eta-\cos\xi},
\end{equation}
where $a$ is the parameter, that at fixed values of pore and granule radii as well as of their center-to- center distance is determined by the relation
 $$a=\frac{\sqrt{[(l-R)^2-R_s^2][(l+R)^2-R_s^2]}}{2\cdot l}.$$
 Pore and granule surfaces in such coordinate system are given by relations 

\begin{equation}\label{eq5}
    \eta_1=\textrm{arsinh} \left(\frac{a}{R}\right), \quad
\eta_2=\textrm{arsinh} \left(\frac{a}{R_s}\right).
\end{equation}
These relations determine values of  $\eta_1$ and $\eta_2$ from pore and granule radii, while  $a$ includes additionally center-to center distance between the pore and the granule. It follows from the above maid suppositions that diffusion fluxes of vacancies and gas atoms onto the pore and granule boundary are determined by stationary diffusion equations as well as by corresponding boundary conditions in the bispherical coordinate system:
\begin{equation}\label{eq6}
 \Delta_{\eta,\xi}c=\frac{\partial}{\partial\eta}\left(\frac{1}{\cosh\eta-\cos\xi}\frac{\partial c}{\partial\eta}\right)+
  \frac{1}{\sin\xi}\frac{\partial}{\partial\xi}\left(\frac{\sin\xi}{\cosh\eta-
  \cos\xi}\frac{\partial c}{\partial\xi}\right)= 0
\end{equation}
\begin{equation}\label{eq7}
  \Delta_{\eta,\xi}c_g =\frac{\partial}{\partial\eta}\left(\frac{1}{\cosh\eta-\cos\xi}\frac{\partial c_g}{\partial\eta}\right)+
  \frac{1}{\sin\xi}\frac{\partial}{\partial\xi}\left(\frac{\sin\xi}{\cosh\eta-
  \cos\xi}\frac{\partial c_g}{\partial\xi}\right)= 0
\end{equation}
\begin{equation}\label{eq8}
c(\eta , \xi)|_{\eta_1}=c_{R}^v, \quad
c(\eta , \xi)|_{\eta_2}=c_{R_s}^v.
\end{equation}
\begin{equation}\label{eq9}
c_g(\eta , \xi)|_{\eta_1}=c_{R}^g, \quad
c_g(\eta , \xi)|_{\eta_2}=c_{R_s}^g.
\end{equation}
 Here we take into account that due to symmetry of the problem, vacancy concentration does not depend on variable $\varphi$. 
General solution of the equations (\ref{eq6})-(\ref{eq7}) is known \cite{17s} and with account of boundary conditions (\ref{eq8})-(\ref{eq9}) it is determined as
\begin{equation}\label{eq10}
c(\eta, \xi ) =
\sqrt{2(\cosh\eta-\cos\xi)}\left\{{c_R^v}\sum_{k=0}^\infty
\frac{\sinh(k+1/2)(\eta-\eta_2)}{\sinh(k+1/2)(\eta_1-\eta_2)}\exp(-(k+1/2)\eta_1)P_k(\cos\xi)-
\right.$$
 $$ \left.- c_{R_s^v}\sum_{k=0}^\infty
\frac{\sinh(k+1/2)(\eta-\eta_1)}{\sinh(k+1/2)(\eta_1-\eta_2)}\exp(-(k+1/2)\eta_2)P_k(\cos\xi)
\right\}\,.
\end{equation}
\begin{equation}\label{eq11}
c_g(\eta, \xi ) =
\sqrt{2(\cosh\eta-\cos\xi)}\left\{{c_R^g}\sum_{k=0}^\infty
\frac{\sinh(k+1/2)(\eta-\eta_2)}{\sinh(k+1/2)(\eta_1-\eta_2)}\exp(-(k+1/2)\eta_1)P_k(\cos\xi)-
\right.$$
 $$ \left.- c_{R_s^g}\sum_{k=0}^\infty
\frac{\sinh(k+1/2)(\eta-\eta_1)}{\sinh(k+1/2)(\eta_1-\eta_2)}\exp(-(k+1/2)\eta_2)P_k(\cos\xi)
\right\}\,.
\end{equation}
Let us note, that here boundary vacancy concentrations $c_R^v, c_{R_s}^v $ are expressed through $\eta_1, \eta_2$ and $a$, and $c_{R_s}$ through $\eta_2$ and $a$. This solution determines stationary vacancy concentration anywhere inside spherical granule of  radius  $R_s$ and outside pore of radius $R$. However, the knowledge of vacancy concentration allows one to find vacancy fluxes onto the pore as well as onto granule boundary at the given positions of granule and pore. These fluxes cause change size and position of pore. With account of this, one can write down the equations for the time change of pore and granule radii as well as of their center-to-center distance. Vacancy flux is determined by the first Fick's low as  
Let us note, that here boundary vacancy concentrations $c_R^v, c_{R_s}^v $ are expressed through $\eta_1, \eta_2$ and $a$, while boundary gas concentrations  $c_R^g, c_{R_s}^g$  are expressed through pressures $P_g$ and $P_0$ correspondingly (see relations (\ref{eq3})).Solutions (\ref{eq10})-(\ref{eq11})  determine stationary concentrations of vacancies and gas atoms   everywhere inside spherical granule of radius $R_s$ and outside the pore of radius $R$. The knowledge of vacancy concentration allows one to find vacancy and atomic fluxes onto the pore as well as onto granule boundary at the given positions of granule and pore. These fluxes cause change size and position of pore. With account of this, one can write down the equations for the time change of pore and granule radii as well as of their center-to-center distance. Vacancy ans atomic fluxes are determined by the first Fick's low as  
\begin{equation}\label{eq12}
\vec{j}_v=-\frac {D}{\omega} \nabla c, \quad \vec{j}_g=-\frac {D_g}{\omega} \nabla c_g
\end{equation}
In the expression (\ref{eq12}) gas atoms flux we also neglected the field of elastic strains.
 Let denote the outer pore surface normal as  $\vec{n}$. Then vacancy flux onto pore surface is determined by scalar product $\vec{n} \cdot \vec{j}|_{\eta=\eta_1}$. Let us write down the expression for vacancy flux onto unit area of pore surface using the expression for gradient in bispherical coordinates \cite{17s}
\begin{equation}\label{eq13}
\vec{n} \cdot \vec{j}_v|_{\eta=\eta_1}=\frac {D}{\omega} \cdot
\frac{\cosh\eta_1-\cos\xi}{a}\frac{\partial
c}{\partial\eta}\left|_{\eta=\eta_1}\right.\,.
\end{equation}
\begin{equation}\label{eq14}
\vec{n} \cdot \vec{j}_g|_{\eta=\eta_1}=-\frac {D_g}{\omega} \cdot
\frac{\cosh\eta_1-\cos\xi}{a}\frac{\partial
c_g}{\partial\eta}\left|_{\eta=\eta_1}\right.\,.
\end{equation}
Similar expression determines vacancy flux onto unit area of granule surface 
\begin{equation}\label{eq15}
\vec{n} \cdot \vec{j}_v|_{\eta=\eta_2}=\frac {D}{\omega} \cdot
\frac{\cosh\eta_2-\cos\xi}{a}\frac{\partial
c}{\partial\eta}\left|_{\eta=\eta_2}\right.\,.
\end{equation}
Here $\vec{n}$ is granule surface normal. Naturally, the total vacancy flux onto pore surface determines the rate of pore volume change: 
\begin{equation}\label{eq16}
\dot{R}=-\frac{\omega}{4\pi R^2} \oint\vec{n}\cdot\vec{j}_v|_{\eta=\eta_1} \, dS  \end{equation}
In the same way one obtains the equation that determines granule radius: 
\begin{equation}\label{eq17}
\dot{R_s}=-\frac{\omega}{4\pi R_s^2}\oint \vec{n}\cdot\vec{j}_v|_{\eta=\eta_2} dS  \end{equation}
Besides this it is necessary to determine the change of gas atoms number in the pore. The rate of this change is determined by gas atoms on the unit area of pore boundary:
\begin{equation}\label{eq18}
\dot{N_g}=\oint\vec{n}\cdot\vec{j_g}|_{\eta=\eta_1} \, dS  \end{equation}
Now let us obtain the equation for the change of pore radius with time. Substituting the exact solution of (\ref{eq10}) into (\ref{eq13}), and performing integration over pore pore surface in (\ref{eq16}) one finds:
\begin{equation}\label{eq19}
    \dot{R}=-\frac{D}{R}\left[\frac{c_R^v}{2}
+\sinh\eta_1\cdot(c_R^v\cdot(\Phi_1+\Phi_2)-2c_{R_s}^v\cdot\Phi_2)\right],
\end{equation}
where functions $\Phi_1$ and $\Phi_2$ are introduced that consist of the sums of exponential series:
\[\Phi_1=\sum_{k=0}^\infty \frac{ e^{-(2k+1)\eta_1}}{e^{(2k+1)(\eta_1-\eta_2)}-1}, \quad \Phi_2=\sum_{k=0}^\infty \frac{ e^{-(2k+1)\eta_2}}{e^{(2k+1)(\eta_1-\eta_2)}-1}\]
The details of deriving these equations are given in the Appendix. Here $\eta_1$ and $\eta_2$ are expressed through pore radii correspondingly to the relations (\ref{eq5}), and $c_R^v$ and $c_{R_s}^v$ are determined via relations (\ref{eq1}), (\ref{eq2}). Thus, the right part of this equation depends nonlinearly on $R$, ${R_s}$ and $l$. In the same way one finds the equation for granule radius change
\begin{equation}\label{eq20}
\dot{R_s}=-\frac{D}{R_s}\left[\frac{c_{R_s}^v}{2}
+\sinh\eta_2\cdot(2c_{R}^v\cdot\Phi_2 - c_{R_s}^v\cdot(\Phi_2+\Phi_3)) \right]\,,
\end{equation}
where function $\Phi_3$ is introduced that is defined as follows:
\[\Phi_3=\sum_{k=0}^\infty \frac{ e^{-(2k+1)(2\eta_2-\eta_1)}}{e^{(2k+1)(\eta_1-\eta_2)}-1}=\sum_{k=0}^\infty \frac{ e^{-(2k+1)\eta_3}}{e^{(2k+1)(\eta_1-\eta_2)}-1}\]
Let us now turn to the equation for the time change of gas atoms number $N_g$ in the pore. Substituting the exact solution (\ref{eq11}) into (\ref{eq14}) and performing integration in (\ref{eq18}),  one obtains the following equation (derivation details are given in the Appendix):
\begin{equation}\label{eq21}
\dot{N_g}=\frac{4\pi D_g}{\omega}\cdot R\cdot\left[-\frac{c_R^g}{2}
+\sinh\eta_1\cdot(-c_R^g\cdot(\Phi_1+\Phi_2)+2c_{R_s}^g\cdot\Phi_2)\right] \end{equation}
In order to obtain a closed set of equations determining granule and pore evolution, one needs to complement the equations (\ref{eq19})-(\ref{eq21}) with one for the rate of changing center-to-center distance between the pore and the granule.  Of course, the displacement rate of vacancy pore relative to granule center is determined by diffusion fluxes of vacancies onto pore surface (see e.g. \cite{7s},\cite{16s} ). In the present case, the displacement rate is determined by relation 
\begin{equation}\label{eq22}
\vec{v}=-\frac{3\omega}{4\pi R^2}\oint \vec{n}(\vec{n}\cdot\vec{j}_v)|_{\eta=\eta_1} dS.
\end{equation}
Using again the exact solution (\ref{eq10}) and performing integration (see Appendix), one obtains: 
\begin{equation}\label{eq23}
\vec{v}=\vec{e_z}\cdot\frac{3 D}{R}\times$$
$$\times
\left[\sinh^2\eta_1 \cdot(c_R^v\cdot
(\widetilde{\Phi}_1+\widetilde{\Phi}_2)-2c_{R_s}^v\cdot \widetilde{\Phi}_2)-\frac{1}{2}\sinh2\eta_1\cdot(c_R^v\cdot(\Phi_1+\Phi_2)-2c_{R_s}^v\cdot \Phi_2)\right]
\end{equation}
Here new functions  $\widetilde{\Phi}_1$ and $\widetilde{\Phi}_2$ are introduced:
\[\widetilde{\Phi}_1=\sum_{k=0}^\infty \frac{(2k+1)e^{-(2k+1)\eta_1}}{e^{(2k+1)(\eta_1-\eta_2)}-1}, \quad \widetilde{\Phi}_2=\sum_{k=0}^\infty \frac{(2k+1)e^{-(2k+1)\eta_2}}{e^{(2k+1)(\eta_1-\eta_2)}-1}\,.\]
Taking into account that displacement rate along  $z$ coincides with $dl/dt$, let us write down the equation in the final form 
\begin{equation}\label{eq24}
   \frac{dl}{dt} =  \frac{3 D}{R}\times$$
$$\times
\left[\sinh^2\eta_1 \cdot(c_R^v\cdot
(\widetilde{\Phi}_1+\widetilde{\Phi}_2)-2c_{R_s}^v\cdot \widetilde{\Phi}_2)-\frac{1}{2}\sinh2\eta_1\cdot(c_R^v\cdot(\Phi_1+\Phi_2)-2c_{R_s}^v\cdot \Phi_2)\right]
\end{equation}
The obtained equation set  (\ref{eq19}), (\ref{eq21}) and (\ref{eq24}) determines completely evolution of the gas-filled pore and the granule. In the limiting case when the gas is absent, $N_g=0$, equations (\ref{eq19}), (\ref{eq21}) and (\ref{eq24}) agree with results of the work \cite{14s}. 

It can be seen from the equations set obtained that the volume of granule material does not change with time since vacancies only carry away emptiness:
\begin{equation}\label{eq25}
R_s(t)^2\dot{R}_s(t)-R(t)^2\dot{R}(t)=0  \end{equation}
The validity of such conservation law is connected closely with current quasi stationary approximation. Vacancy  fluxes, that come out from the pore and from the granule are balanced with each other. Thus, the volumes of the pore and of the granule are connected with each other by an easy relation
\begin{equation}\label{eq26}
  R_s(t)^3 ={V+R(t)^3},
\end{equation}
where $V=R_s(0)^3 -R(0)^3$ is initial volume of granule material (multiplier  $4\pi /3$ is omitted for convenience). 
The existence of conservation low (\ref{eq26}) enables us to reduce the number of unknown quantities. As a result, we obtain Cauchy problem for the system of three  differential equations for $R$, $l$ and $N_g$, whose solution describes evolution of gas-filled pore inside the granule:

\begin{equation}\label{eq27}
\begin{cases}
\frac{d l}{d t}=\frac{3Dc_V}{R}\cdot\exp\left(\frac{2\gamma\omega}{kTR}-\frac{3\omega N_g}{4\pi R^3}\right) \cdot \left[\frac{a^2}{R^2} \cdot (\widetilde \Phi _1  + \widetilde \Phi _2 ) - \frac{a}{R} \cdot \sqrt {1 + \frac{a^2}{R^2}}  \cdot (\Phi _1  + \Phi _2 )\right]-\\
-\frac{6Dc_V}{R}\cdot\exp\left(-\frac{2\gamma\omega}{kTR_{s}}-\frac{P_0\omega}{kT}\right)\cdot \left[\frac{a^2}{R^2} \cdot \widetilde \Phi _2  - \frac{a}{R} \cdot \sqrt {1 + \frac{a^2}{R^2}}  \cdot \Phi _2 \right],\\
\frac{d R}{d t}=-\frac{Dc_V}{R}\cdot\exp\left(\frac{2\gamma\omega}{kTR}-\frac{3\omega N_g}{4\pi R^3}\right) \cdot \left[\frac{1}{2} + \frac{a}{R} \cdot (\Phi _1  + \Phi _2 )\right]+\\
+\frac{2Dc_V}{R}\cdot\exp\left(-\frac{2\gamma\omega}{kTR_{s}}-\frac{P_0\omega}{kT}\right)\cdot \frac{a}{R} \cdot \Phi _2,\\
\frac{d N_g}{d t}=-\frac{3D_gN_g\kappa_H kT}{\omega R^2 }\cdot\left[\frac{1}{2} + \frac{a}{R} \cdot (\Phi _1  + \Phi _2 )\right]+\\
+\frac{8\pi D_g }{\omega}\cdot \kappa_H P_0 \cdot a \Phi _2,\\
  R_s =\sqrt[3]{V+R^3},\; P_0=\textrm{const},  \\
  R|_{t=0}=R(0), R_s|_{t=0}=R_s(0),\\
  l|_{t=0}=l(0),\\
	N_g|_{t=0}=N(0).
\end{cases}
\end{equation}
For the sake of convenience, let us introduce dimensionless variables  with characteristic length $R_0 = R(0)$ (that is pore radius at the initial time moment $t = 0$) and characteristic time $t_g = \omega R_0^2/3D_g\kappa_H kT$. Let us now go over to the following dimensionless variables: 
\[ r = \frac{R}{R_0},\quad r_s = \frac{R_{s}}{R_0},\quad L = \frac{l}{R_0},\quad {\tau} = \frac{t}{{t_g }},\quad t_g=\frac{\omega R_0^2}{3D_g\kappa_H kT},  \quad \alpha = \frac{a}{R_0},\quad  \frac{{2\gamma \omega }}{{kTR}} = \frac{A}{{r}}, \]
\begin{equation}\label{eq28}
\frac{{2\gamma \omega }}{{kTR_s}} = \frac{A}{{r_s}},\quad A = \frac{{2\gamma \omega }}{{kTR_0}},\quad \frac{3\omega N_g}{4\pi R^3} = \frac{B}{r^3}\cdot n, \quad B=\frac{3\omega N_0}{4\pi R_0^3}, \quad n=\frac{N_g}{N_0},\end{equation}
\[\frac{P_0\omega}{kT}=p_0,\quad q=\frac{t_V}{t_g}=\frac{3D_g \kappa_H kT}{Dc_V\omega}, \quad t_V=\frac{R_0^2}{Dc_V}. \]

Using these variables, one can rewrite the equation set  (\ref{eq27}) in dimensionless form: 
\begin{equation}\label{eq29}
\begin{cases}
  \frac{d L}{d \tau}=\frac{3\exp\left(\frac{A}{r}-\frac{B}{r^3}\cdot n\right)}{qr} \cdot \left[\frac{\alpha^2}{r^2} \cdot (\widetilde \Phi_1  + \widetilde \Phi_2 ) - \frac{\alpha}{r} \cdot \sqrt {1 + \frac{\alpha^2}{r^2}}  \cdot (\Phi_1  + \Phi_2 )\right]-\\
-\frac{6\exp\left(-\frac{A}{r_s}-p_0\right)}{qr}\cdot \left[\frac{\alpha^2}{r^2} \cdot \widetilde \Phi_2  - \frac{\alpha}{r} \cdot \sqrt {1 + \frac{\alpha^2}{r^2}}  \cdot \Phi_2 \right],\\
\frac{d r}{d \tau}=-\frac{\exp\left(\frac{A}{r}-\frac{B}{r^3}\cdot n\right)}{qr} \cdot \left[\frac{1}{2} + \frac{\alpha}{r} \cdot (\Phi_1  + \Phi_2 )\right]+\frac{2\exp\left(-\frac{A}{r_s}-p_0\right)}{qr}\cdot \frac{\alpha}{r} \cdot \Phi_2,\\
\frac{d n}{d \tau}=-\frac{n}{r^2} \cdot \left[\frac{1}{2} + \frac{\alpha}{r} \cdot (\Phi_1  + \Phi_2 )\right]+2\alpha\cdot \frac{p_0}{B} \cdot \Phi _2,\\
r_s=\sqrt[3]{V+r^3},\; p_0=\textrm{const},\; N_0=N_g|_{\tau =0}=\textrm{const}, \\
  r|_{\tau =0}=1, r_s|_{\tau =0}=r_{s0}, \\
  L|_{\tau =0}=\frac{l(0)}{R(0)},\\
	n|_{\tau =0}=1.
\end{cases}
\end{equation}
In the limiting case when the gas is absent, equation set (\ref{eq29}) turns to that obtained in the work \cite{14s} for the behavior of vacancy pore.

From the analysis of this equation set already follows the principal difference in the behavior of gas-filled pore inside the granule as compared to its behavior in macroscopic bodies. One can check, that nonlinear equation set (\ref{eq29}) has no fixed points. Physically, this means impossibility of the stabilization of a gas-filled pore. Indeed, let us suppose that the number of gas atoms in the pore became stabilized and stopped changing. This means, that gas fluxes into the pore are absent, but this is possible only when equilibrium pressure $P_0$ is reached in the pore. However, on this conditions vacancy fluxes do not vanish because of inequalities of equilibrium concentrations of vacancies near the boundaries. The last means that pore radius will change and? correspondingly, equality to zero of the right part of the equation determining pore radius can not be realized. This, in its turn, signifies the absence of fixed points.

Let us now consider the alternative case when vacancy fluxes ceased and pore stopped changing its size size. This means that boundary concentrations of vacancies at the boundaries are equal to each other (see relations (\ref{eq1}) and (\ref{eq2})). This is possible if conditions $ 2 \omega \gamma (1/R + 1/R_s)=P_0 -P_g$ are met. However, the performance of these conditions does not mean ceasing of gas flux into the pore because of different equilibrium concentrations of gas at pore and granule boundaries (see relations (\ref{eq3})). This signifies that pressure $P_g$ in the pore will grow and the above conditions will be violated.

In a simple case, when pore displacement relative to granule center $L(\tau)\approx const$ is small, the above said can be verified via numerical methods of qualitative analysis of the set of two equations. It is worth to notice that the case described is realistic enough. Indeed, with the use of simple geometrical considerations and the estimate of  vacancy fluxes asymmetry  arising from pore displacement, one can estimate characteristic value of pore displacement as $ \Delta l \approx R_0 l_0/(R_s -R_0)$. This estimate demonstrates the smallness of pore displacement in the granule.

Thus, the above said signifies the impossibility of stabilization of  gas-filled pore in bounded particles of limited size under considered conditions.

\section{Asymptotic evolution modes}

 Let us investigate the equation set (\ref{eq29}) in various limiting cases. This will help us to understand more clearly pore evolution modes. These modes are determined by combinations of three dimensionless values: $R/R_s$, $l/R_s$  and  $R/l$. Let us suppose that the pore is situated at the distance $l$ from the granule center and its radius equals to $R$. These values are limited by purely geometrical inequality:
\begin{equation}\label{eq1A}
    R/R_s+l/R_s < 1
\end{equation}
Physically this means that the pore is situated inside the granule and does not touch its boundary. Besides this, the value $\delta=R/R_s < 1$   is always smaller the unity, since pore size cannot exceed that of the granule. Let us introduce conventional classification of pores as "small" and "large" ones according to the relation of their size to that of the granule. Let us consider in more details the asymptotic modes that can be realized.

\subsection{Small pores}

Let us define pores as small one if the condition $R/R_s \ll 1$ is obeyed. Let us, similar to the work \cite{14s}, consider various possibilities of pore position relative to the  granule center. There can be three possibilities. First one is that distance from a small pore to the granule center is large as compared to granule radius.:
 \begin{equation}\label{eq2A}
  \quad R/R_s \ll 1, \quad l/R_s \ll 1,\quad R/l \ll 1
\end{equation}
Second possibility is that a small pore is situated close to the granule center:
\begin{equation}\label{eq3A}
   \quad R/R_s \ll 1, \quad l/R_s \ll 1,\quad R/l \gg 1
\end{equation}
Finally, there exist a third possibility that small pores are situated at significant distance from the granule center that is comparable with granule size. In this case, pore is situated close to granule boundary. 
\begin{equation}\label{eq4A}
  \quad R/R_s \ll 1, \quad l/R_s \simeq 1, \quad R/l \ll 1.
\end{equation}
 It can be seen easily that evolution of small pores 
$\frac{R(0)}{R_s(0)} \ll 1$ cannot be accompanied by a significant change of granule dimensions. Indeed, using the relation (\ref{eq19})(26), one can estimate an order of granule size change during the evolution. According to  (\ref{eq26}) this change can be written down in the form: 
\[\frac{R_s(t)}{R_s(0)}=\sqrt[3]{1-\frac{R(0)^3}{R_s(0)^3}+\frac{R(t)^3}{R_s(0)^3}} \simeq 1-\frac{1}{3}\left(\frac{R(0)^3}{R_s(0)^3}-\frac{R(t)^3}{R_s(0)^3}\right)  \]
Hence, within small pore approximation, granule size does not change  $R_s(t) \approx R_s(0)=R_{0s}$ up to third order of smallness $\frac{R(0)^3}{R_s(0)^3} $, $\frac{R(t)^3}{R_s(0)^3} $. Then, neglecting granule radius change, the equation system (\ref{eq29}) takes on the following form:
\begin{equation}\label{eq5A}
\begin{cases}
  \frac{d L}{d \tau}=\frac{3\exp\left(\frac{A}{r}-\frac{B}{r^3}\cdot n\right)}{qr} \cdot \left[\frac{\alpha^2}{r^2} \cdot (\widetilde \Phi_1  + \widetilde \Phi_2 ) - \frac{\alpha}{r} \cdot \sqrt {1 + \frac{\alpha^2}{r^2}}  \cdot (\Phi_1  + \Phi_2 )\right]-\\
-\frac{6\exp\left(-\frac{A}{r_{s0}}-p_0\right)}{qr}\cdot \left[\frac{\alpha^2}{r^2} \cdot \widetilde \Phi_2  - \frac{\alpha}{r} \cdot \sqrt {1 + \frac{\alpha^2}{r^2}}  \cdot \Phi_2 \right],\\
\frac{d r}{d \tau}=-\frac{\exp\left(\frac{A}{r}-\frac{B}{r^3}\cdot n\right)}{qr} \cdot \left[\frac{1}{2} + \frac{\alpha}{r} \cdot (\Phi_1  + \Phi_2 )\right]+\frac{2\exp\left(-\frac{A}{r_{s0}}-p_0\right)}{qr}\cdot \frac{\alpha}{r} \cdot \Phi_2,\\
\frac{d n}{d \tau}=-\frac{n}{r^2} \cdot \left[\frac{1}{2} + \frac{\alpha}{r} \cdot (\Phi_1  + \Phi_2 )\right]+2\alpha\cdot \frac{p_0}{B} \cdot \Phi _2,\\
 p_0=\textrm{const}, \\
  r|_{\tau =0}=1, r_s|_{\tau =0}=r_{s0}, \\
  L|_{\tau =0}=\frac{l(0)}{R(0)},\\
	n|_{\tau =0}=1.
\end{cases}
\end{equation}
Here parameter $\alpha$ is of the following form:
\[\alpha = \frac{r_s^2}{2L}\sqrt{1+\left(\frac{L^2}{r_s^2}-\frac{r^2}{r_s^2}\right)^2-2\left(\frac{L^2}{r_s^2}+\frac{r^2}{r_s^2}\right)}.\]
Thus, the behavior of small pore is described by nonlinear equation set (\ref{eq5A}).

\subsubsection{The pore far from granule center }

Let us now consider asymptotic case (\ref{eq2A}). With the account that $L \gg r$, in equations (\ref{eq5A}) the expression  for the parameter $\alpha$ is simplified
\begin{equation}\label{eq6A}
\alpha \approx \frac{r_{s0}^2}{2L}\sqrt{\left(1-\frac{L^2}{r_{s0}^2}\right)^2}= \frac{r_{s0}^2}{{2L}}\left(1-\frac{L^2}{r_{s0}^2}\right),\end{equation}
while bispherical coordinates $\eta_{1,2}$, determined by (\ref{eq6}), are, correspondingly:
\begin{equation}\label{eq7A}
 \eta _1  = \textrm{arsinh} \left( {\frac{r_{s0}^2}{{2rL}}}\left(1-\frac{L^2}{r_{s0}^2}\right) \right),\;\eta _2  = \textrm{arsinh} \left( \frac{r_{s0}}{2L} \left(1-\frac{L^2}{r_{s0}^2}\right) \right).\end{equation}
Since $\frac{\sinh\eta_1}{\sinh\eta_2}=\frac{r_{s0}}{r} \gg 1$, then $\eta_1  \gg \eta_2$. In this case series sums are estimated by the following expression:
\begin{equation}\label{eq8A}
 \Phi_1 \approx \frac{1}{2\sinh2\eta_1},\; \Phi_2 \approx \frac{1}{2\sinh\eta_1},\;  \widetilde{\Phi}_1 \approx \frac{1+2\sinh^2\eta_1}{8\sinh^2\eta_1\cosh^2\eta_1},\; \widetilde{\Phi}_2 \approx \frac{\cosh\eta_1}{2\sinh^2\eta_1},$$
 $$ \sinh\eta_1=\frac{\alpha}{r},\quad \cosh\eta_1=\sqrt{1+\frac{\alpha^2}{r^2}}. \end{equation}
Substitution of (\ref{eq6A})-(\ref{eq8A}) into equation set  (\ref{eq5A}) yields the following equation set:
\begin{equation}\label{eq9A}
\begin{cases}
\frac{d L}{d \tau}=-\frac{3}{2}\cdot\exp\left(\frac{A}{r}-\frac{B}{r^3}\cdot n\right)\cdot\frac{r \left(\frac{L^2}{r_{s0}^2}\right)}{qr_{s0}^2}  \\
\frac{d r}{d \tau}=-\frac{\exp\left(\frac{A}{r}-\frac{B}{r^3}\cdot n\right)}{qr} \cdot \left[1 + \frac{r}{2L}\cdot\left(\frac{L^2}{r_{s0}^2}\right) \right]+\frac{\exp\left(-\frac{A}{r_{s0}}-p_0\right)}{qr}  \\
\frac{d n}{d \tau}=-\frac{n}{r^2} \cdot \left[1 + \frac{r}{2L}\cdot\left(\frac{L^2}{r_{s0}^2}\right) \right]+r\cdot \frac{p_0}{B}  \\
\end{cases}
\end{equation}
Equations (\ref{eq9A}) are written down up to  $L^2/r_{s0}^2$ terms. The right part of the equation determining distance from granule center is negative and is negative and has square order of smallness. This signifies that pore is always moving slowly towards the granule center. It is easy to prove  that if  gas pressure in the pore $P_g$ is smaller then outer pressure $P_0$,  the sign of the right part of the second equation is negative. Correspondingly, pore size under these conditions decreases monotonously while pore is moving towards the granule center. If  gas pressure in the pore $P_g$ exceeds outer one, the sign of the right part is positive that signifies that pore is at the initial stage of growth. However, as pore radius continues to increase, the sign of the right part becomes negative and the pore starts to diminish its size.

It can be seen from the right part of the third equation of the set (\ref{eq9A}), that in the beginning of pore evolution, the dominating term is proportional to outer pressure $P_0$. Gas pressure inside the pore is smaller then $P_0$. This leads to the increase the number of gas atoms in the pore until pore size decreases down to certain value at which  gas pressure in the  pore reaches $P_0$. After that the negative term is determined by the term of the second order of smallness $\sim(-L^2/r_{s0}^2)$. As result, the number gas atom in the pore starts to decrease.

Let us turn now to the numerical analysis of the complete equation set (\ref{eq5A}) under condition that outer pressure $P_0$ exceeds initial gas pressure  $P_g(0)=\frac{N_g(0)kT}{\frac{4\pi}{3}R_0^3}$ in a pore: $P_0>P_g(0)$.

Let us conduct all calculations for the same initial conditions: $r|_{\tau =0}=1$, $r_s|_{\tau =0}=100$, $L|_{\tau =0}=10$, $n|_{\tau =0}=1$, $A=10^{-3}$, $B=2.5\cdot 10^{-3} (N_g(0)=1.04 \cdot 10^9)$ that correspond to small pore far from granule center. Let us choose outer pressure $P_0= 10^8$ dyn/cm${^2} (P_0>P_g(0))$  or in dimensionless 
\begin{figure}
  \centering
  \includegraphics[width=5.5 cm, height=5.5 cm]{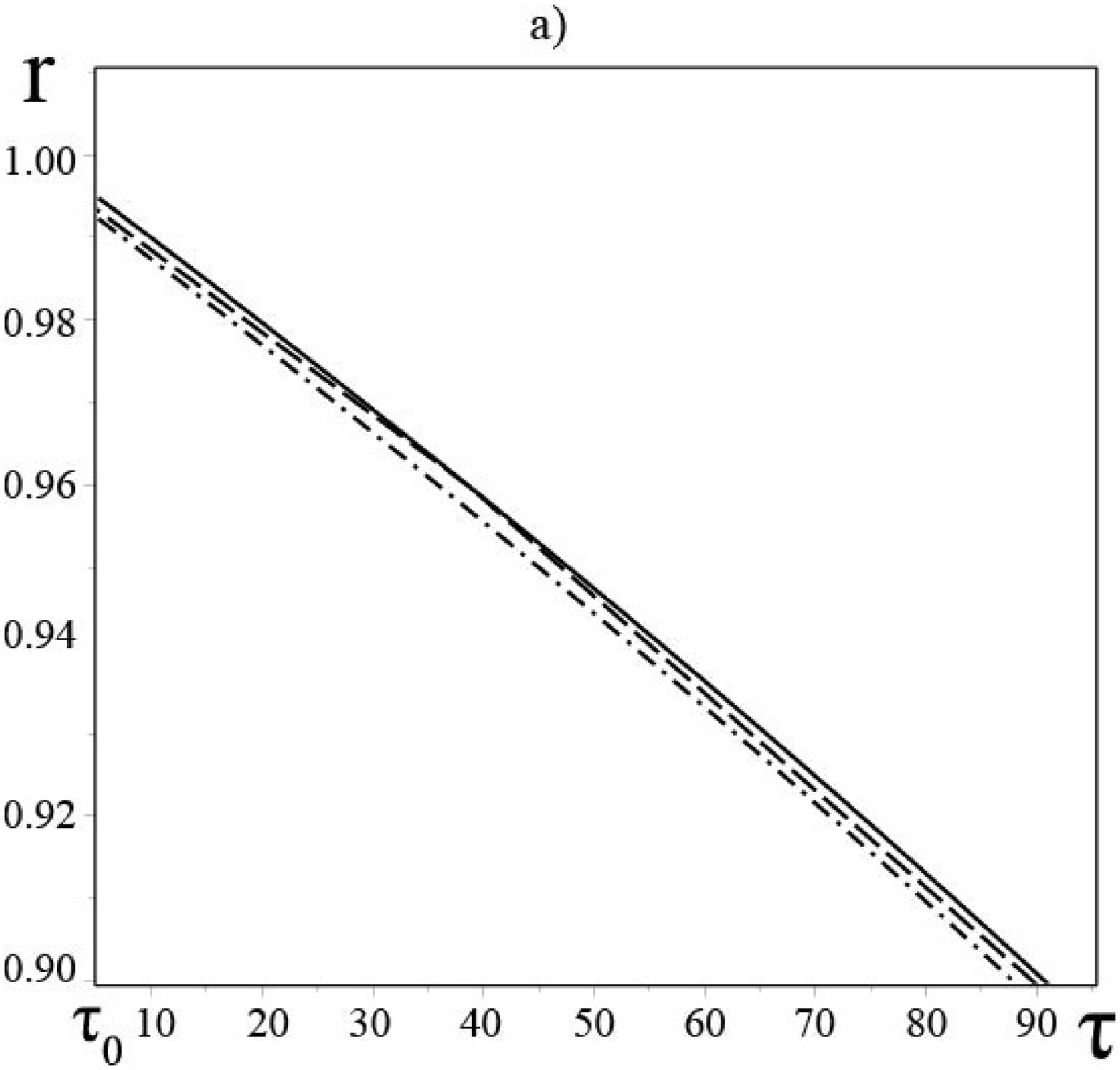}
	\includegraphics[width=5.5 cm, height=5.5 cm]{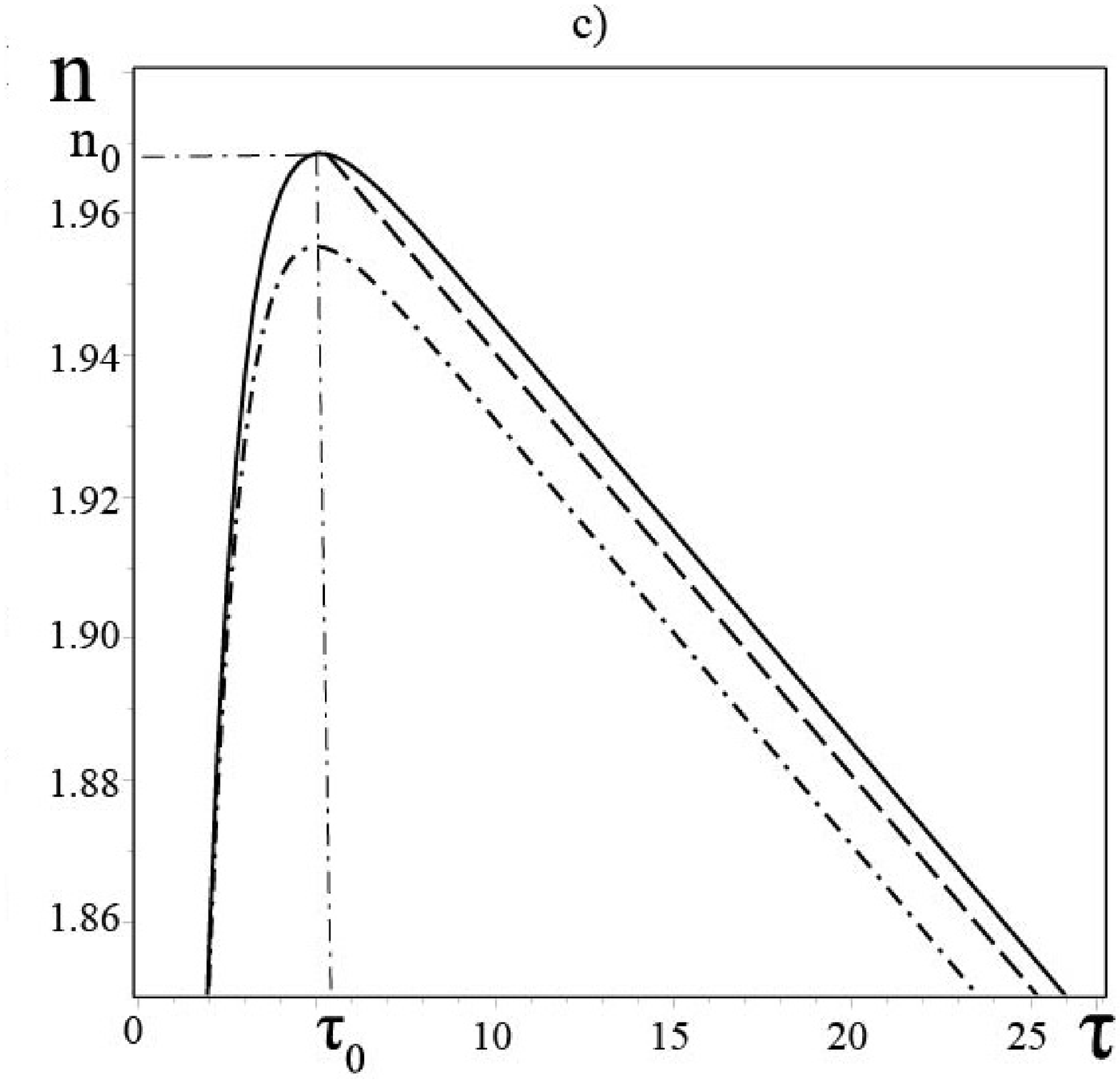}
	\includegraphics[width=5.6 cm, height=5.5 cm]{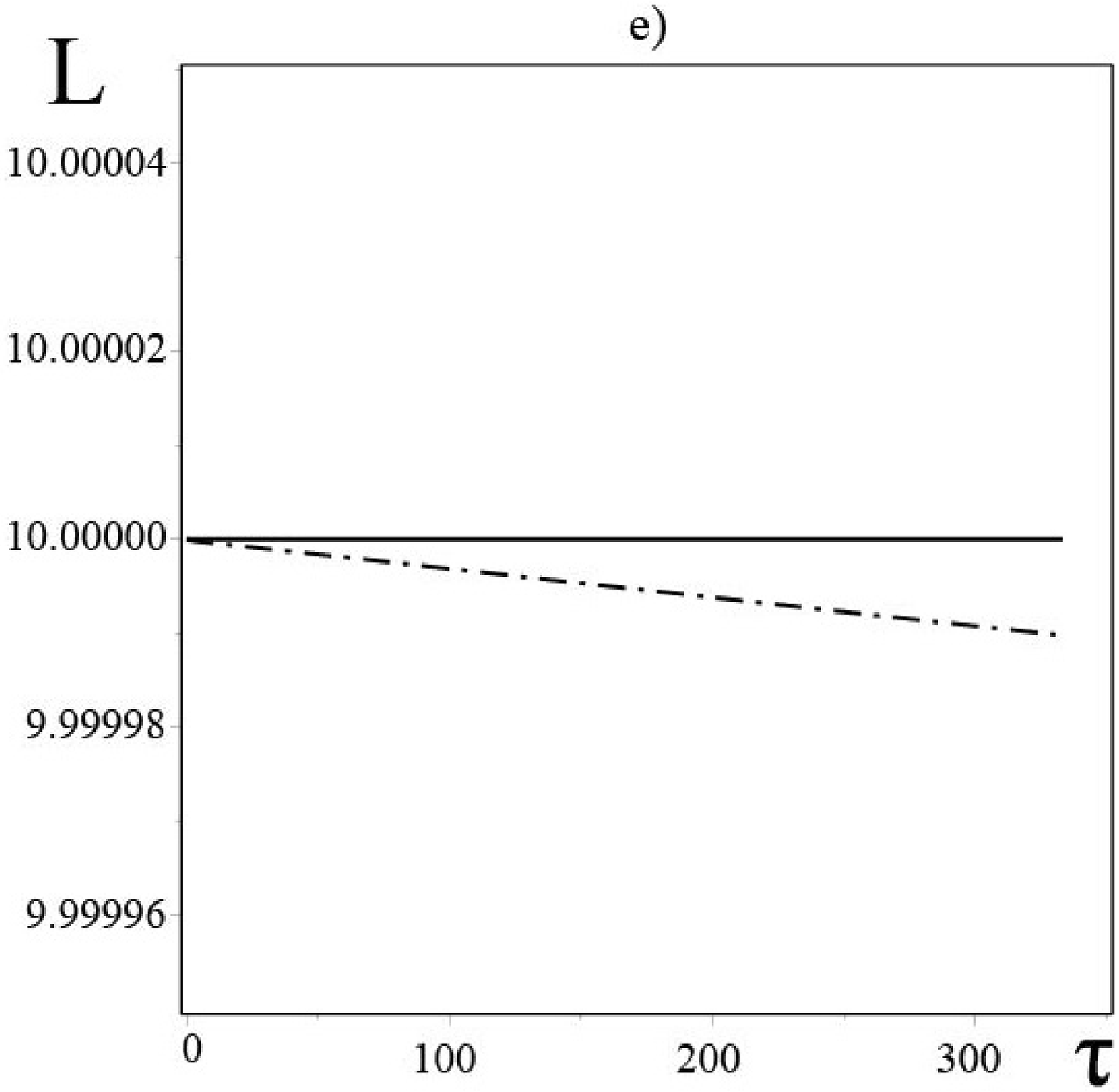}\\
	\includegraphics[width=5.5 cm, height=5.5 cm]{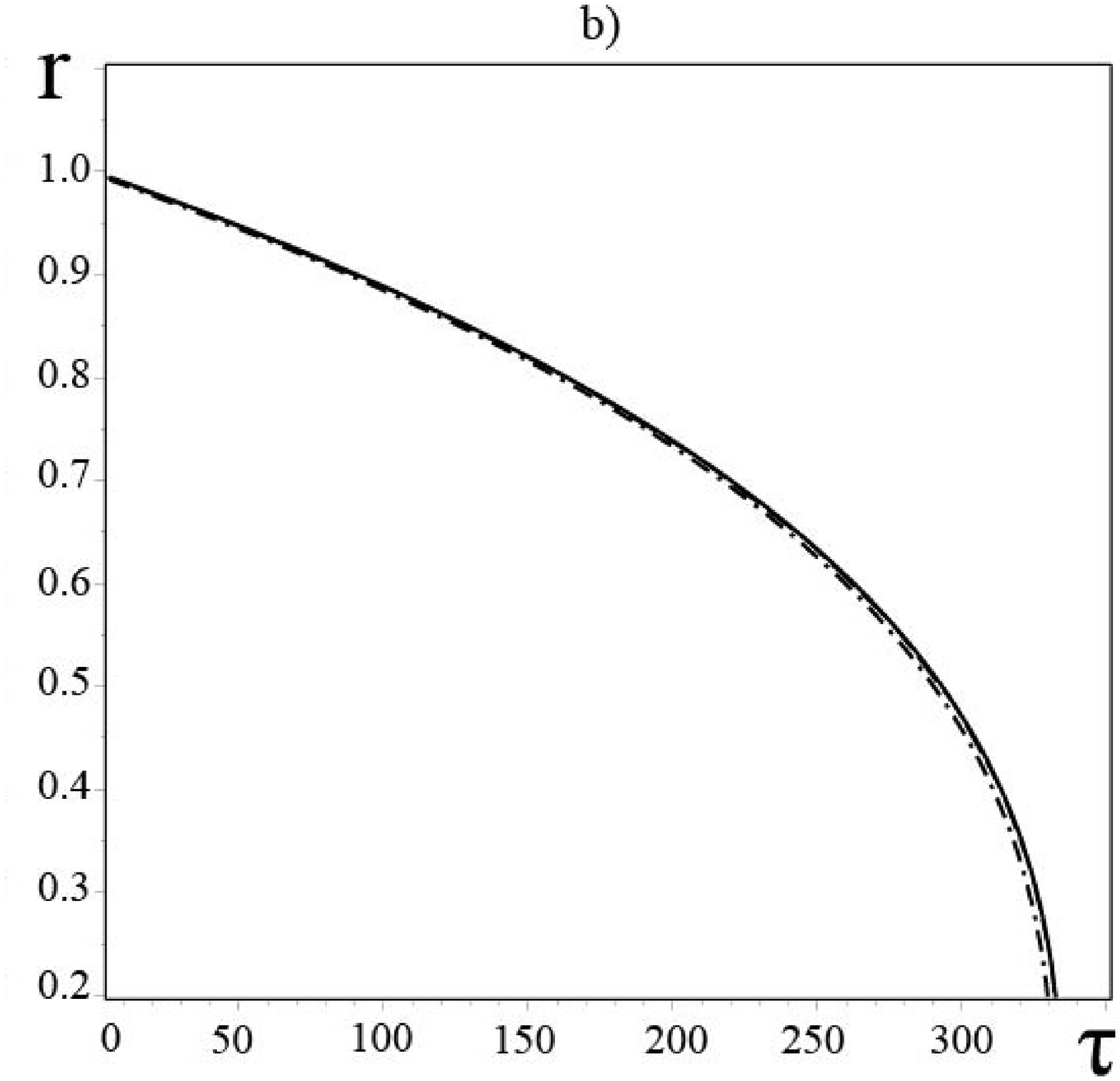}
	\includegraphics[width=5.5 cm, height=5.5 cm]{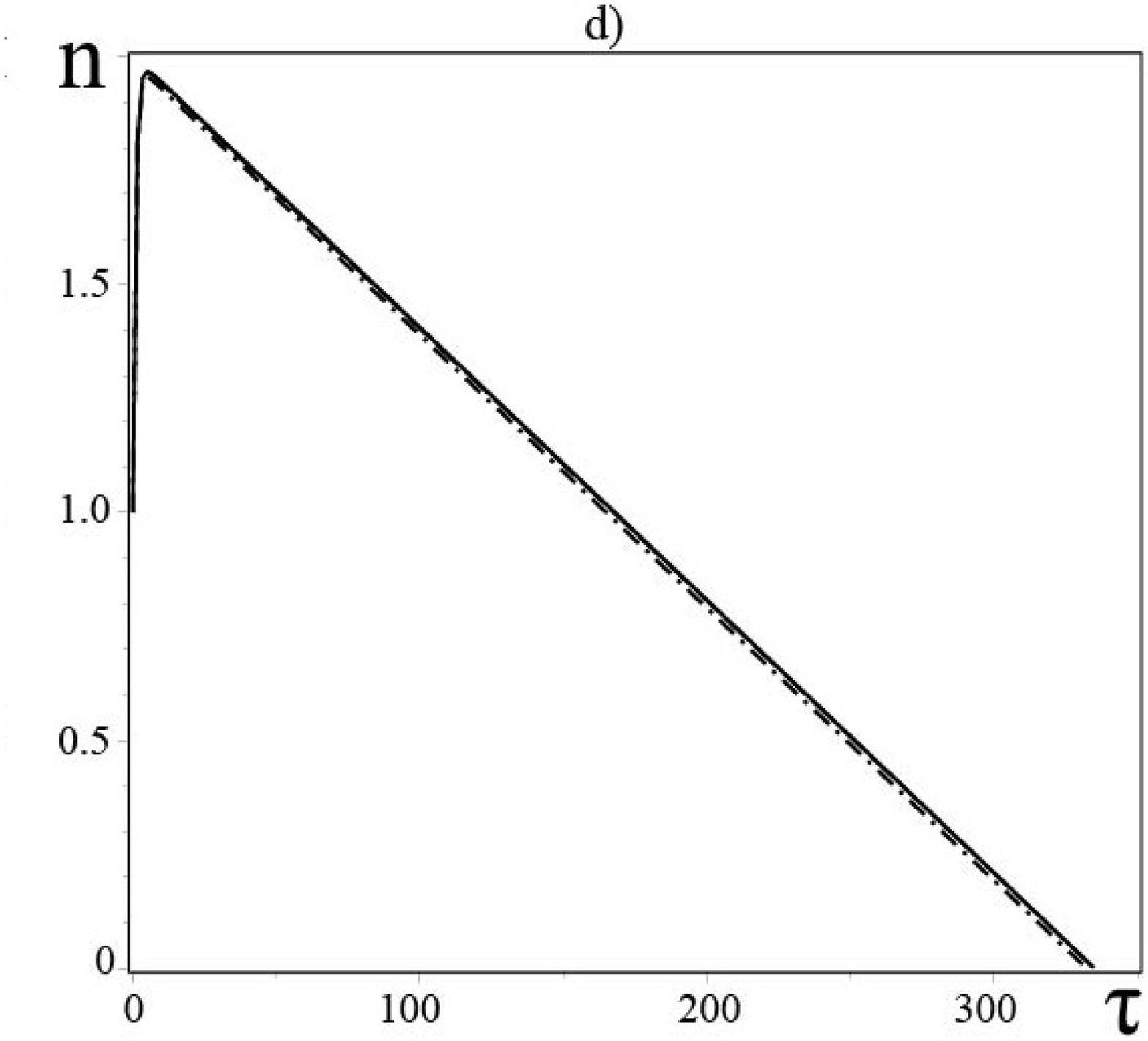}
	\includegraphics[width=5.6 cm, height=5.5 cm]{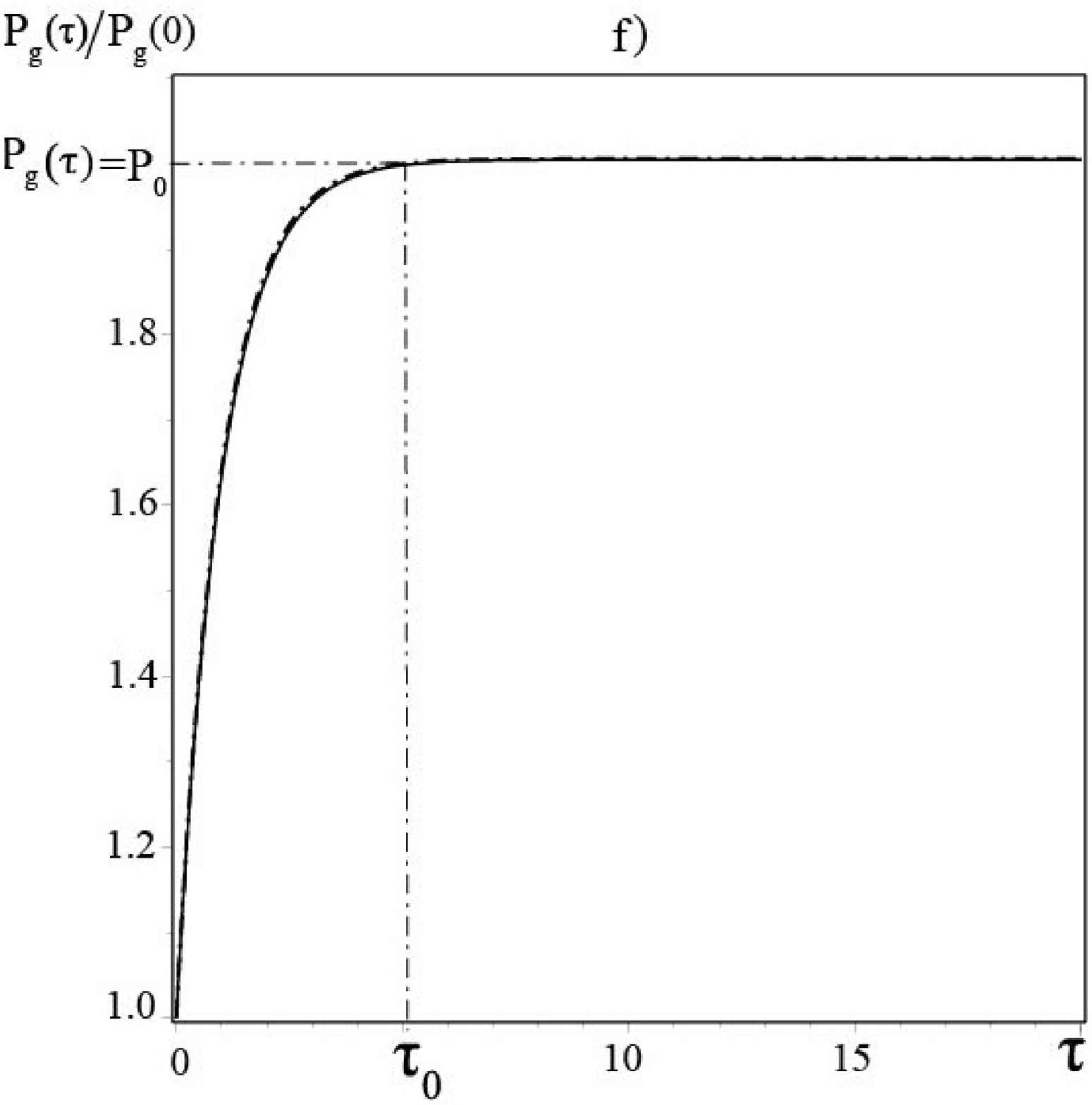}\\
  \caption{a)-b) plots of the dependence of pore radius $r$ on time $\tau$;  c)-d) plots of the change of gas atoms number $n$ in the pore with time $\tau$; e) -- plots of  the change of distance $L$ with time $\tau$; f) -- plots of the change of gas pressure $P_g(\tau)/P_g(0)$ in the pore on time $\tau$. Dash-and-dot line in the plots corresponds to the solutions of the  complete equation set (\ref{eq5A}), solid line relates to numerical solutions of approximate equation set  (\ref{eq10aa}), dashed line shows analytical solutions of approximate equation set (\ref{eq10aa}). All solutions are obtained for initial conditions $r|_{\tau=0}=1$, $r_s|_{\tau=0}=100$, $L|_{\tau=0}=10$, $n|_{t=0}=1$, $A=10^{-3}$, $p_0 = 5\cdot 10^{-3}$, $q=1$, $B=2.5\cdot 10^{-3}$.}
    \label{fg2}
\end{figure}
units $p_0=5\cdot10^{-3}$. Dimensionless parameter $q=1$,  while dimensionless parameter $A$ during all calculations select equal to $A=10^{-3}$, that corresponds to the following: surface energy $\gamma=10^3$ erg/sm${^2}$, specific volume of the lattice $\omega=10^{-23}$ sm${^3}$, initial pore radius $R(0)=10^{-4}$ sm, Boltzman constant $k=1.38\cdot10^{-16}$ erg/K  and material temperature $T=1450^{\circ}\textrm{K}$.

From the numerical solutions of the complete equation set (\ref{eq5A}), shown in Fig.\ref{fg2}, it is easy to notice several important peculiarities of pore behavior  under chosen conditions. First of all let us consider behavior gas pressure in a pore (see. Fig.\ref{fg2}f). After some period of time $\tau_0$  gas pressure  in a pore reaches the value of outer pressure and, since then, remains the same until pore disappears. Besides that, it is demonstrated in Fig.\ref{fg2} that the number of gas atoms in the pore decreases linearly with time after reaching its maximum value at time moment $\tau_0$. Taking into account that gas state equation connects pressure in the pore and pore volume, one can see that after the lapse of time $\tau_0$ pore volume also diminishes linearly with time. Pore position change is small as it can be seen from Fig.\ref{fg2}e. Thus, in this regime,  pore evolution occurs for most of the time at constant gas pressure in the pore equal to the outer pressure $P_0$.

Taking into account this observation, one can describe pore behavior by simple analytical relations via solving approximate equation set (\ref{eq9A}) for the time $\tau \geq \tau_0$. Let us consider the regime when the gas pressure $P_g$ equal to outer pressure $P_0$ has stabilized in the pore that is equivalent to satisfying conditions
\begin{equation}\label{eq9aa}
 n(\tau)/r^3(\tau)=p_0/B
\end{equation}
Beside this, let us suppose that surface energies of the pore and granule boundaries as well as gas filled pore size are small. 
\[   \frac{A}{r} \ll 1, \quad \frac{A}{r_s} \ll 1, \quad \frac{r}{r_s} \ll 1, \quad \frac{L}{r_s} \ll 1,  \quad   \frac{r}{r_{s}}\frac{L}{r_{s}} \ll \frac{A}{r}\]
Then equation set (\ref{eq9A}) simplifies and takes on the following form:
\begin{equation}\label{eq10aa}
\begin{cases}
L(\tau)\approx L|_{\tau=0}=L_0 \\
\frac{d r}{d \tau}=-\frac{A}{qr^2}\cdot\left[1 + \frac{r L_0}{2r_{s0}^2} \right]\cdot e^{-p_0}   \\
\frac{d n}{d \tau}=-\frac{n}{r^2} \cdot \left[1 + \frac{r L_0}{2r_{s0}^2} \right]+r\cdot \frac{p_0}{B}  \\
\end{cases}
\end{equation}
As it can be seen from equation set (\ref{eq10aa}), equation for pore radius evolution $r(\tau)$ becomes independent on the rest so that it can be solved analytically. Integrating this equations results in analytical solution in the form
\begin{equation}\label{eq11aa}
\tau-\tau_0=-\frac{q}{A} \cdot e^{p_0}\cdot\left(f(\tau)-f(\tau_0)\right),
\end{equation}
where the designation is introduced for the function $$f(\tau)=\frac{r^3(\tau)}{3}-\widetilde \alpha \cdot \frac{r^4(\tau)}{4}, \quad \widetilde\alpha=\frac{L_0}{2r_{s0}^2}.$$
The explicit form of the solution in zero order over $\frac{L_0}{2r_{s0}^2}$ can be obtained easily
\begin{equation}\label{12bb}
    r=\left(r^3(\tau_0) -\frac{3A}{q}\cdot e^{-p_0}(\tau -\tau_0) \right)^{\frac{1}{3}}
\end{equation}
The correction proportional to $\frac{L_0}{2r_{s0}^2}$ be obtained easily via implementing iteration method. Let us note that, according to zero order approximation, pore volume diminishes linearly with time.

Let us now compare solutions of the complete equation set (\ref{eq5A}) with numerical and analytical solutions of approximate equation set (\ref{eq10aa}). In Fig. \ref{fg2}a, for the convenience of tracing corresponding solutions (as in Fig. \ref{fg2}c, Fig. \ref{fg2}e), the behavior of pore radius at small  times is shown. The good agreement is demonstrated  of numerical solutions for the evolution of pore radius of the approximate equation set  (\ref{eq10aa}) and complete equation set (\ref{eq5A}), at the time up to $\tau_0 \leq \tau \leq 50$. In Fig. \ref{fg2}$\textrm{b}$ evolution pore radius $r$ is shown until the <<complete>> pore dissolving.  The plots in Fig.\ref{fg2}c,\ref{fg2}d demonstrate evolution of the number of gas atoms in the pore. In Fig. \ref{fg2} the process is demonstrated of filling the pore by gas until the number of gas atoms reaches certain value $n_0$ at time moment $\tau_0$. During time $\tau \geq\tau_0$ from the beginning of the process, the solutions of approximate (\ref{eq10aa}) and complete (\ref{eq5A}) equation set practically coincide. Very good agreement can be also seen of analytical (\ref{eq11aa}) and numerical solutions of equation sets (\ref{eq5A}) and (\ref{eq10aa}). After the number of atoms in the pore $n_0$ is reached, gas pressure in the pore $P_g$ stabilizes at the value equal to the outer pressure $P_0$. In other words, quick adjustment occurs of gas atoms number to the corresponding pore volume. Time dependence of the distance between  pore and granule centers is shown in Fig. \ref{fg2}e. The displacement of the pore towards granule center in both cases is very small.  In plot \ref{fg2}$\textrm{f}$ evolution of gas pressure $P_g(\tau)/P_g(0)=n(\tau)/r^3(\tau)$ in the pore is shown. The solutions of approximate (\ref{eq10aa}) and complete (\ref{eq5A}) equation set for the evolution of  pressure $P_g(\tau)/P_g(0)=n(\tau)/r^3(\tau)$ coincide completely.
Thus, on this stage pore behavior is described by simple analytical relations (\ref{12bb}) and the number of gas atoms in the pore equals to
\begin{equation}\label{13bb}
    n(\tau)=\frac{p_0}{B} \left(r^3(\tau_0) -\frac{3A}{q}\cdot e^{-p_0}(\tau -\tau_0) \right)
\end{equation}
 This explains well  linear decrease of $n(\tau)$ with time $\tau \geq \tau_0$ that can be noticed in in Fig. \ref{fg2}c,d. From analytical solution (\ref{12bb}) follows an estimate of the lifetime of gas-filled pore in bounded matrix particle:  
\[\Delta \tau \approx \frac{qr^3(\tau_0)}{3A}e^{p_0}\]
 However, let us take into account that the time $\tau_0$ of establishing constant pressure regime in the pore is much smaller then pore lifetime $\Delta \tau$ and, consequently, $r(\tau_0) \approx r(0) =1$. Thus, let us estimate it from the relation (\ref{12bb}) as
\begin{equation}\label{14bb}
    \Delta \tau \approx \frac{q}{3A}e^{p_0}
\end{equation}
This relation determines the dependence of pore lifetime from basic parameters on granule material as well as on granule size. For the parameters of the numerical count, this relation gives $\Delta \tau \approx 330$, that coincides well with results shown in Fig.\ref{fg2}.

\subsubsection{Pore close to granule center}

Let us now turn to the case determined by inequalities (\ref{eq3A}). This inequalities agree with geometrical condition  $R/R_s+l/R_s \leqslant 1$. With account of (\ref{eq3A}), it is easy to find expression for parameter $\alpha$ and byspherical coordinates values $\eta_{1,2}$:
\begin{equation}\label{eq10A} \alpha \approx \frac{r_{s0}^2}{2L}\left(1- \frac{r^2}{r_{s0}^2}\right),\; \eta _1  \approx \textrm{arsinh} \left( {\frac{r_{s0}^2}{{2rL}}}\left(1-\frac{r^2}{r_{s0}^2}\right) \right),\;\eta _2  \approx \textrm{arsinh} \left( \frac{r_{s0}}{2L} \left(1-\frac{r^2}{r_{s0}^2}\right) \right)
\end{equation}
From the above said follows, that, for small pores, the relation $\eta_1 \gg \eta_2 $ is valid. Using the estimate of series sums in (\ref{eq8A}), one obtains approximate equations fore pore evolution for this case:
\begin{equation}\label{eq11A}
\begin{cases}
\frac{d L}{d \tau}=-\frac{3}{2}\cdot\exp\left(\frac{A}{r}-\frac{B}{r^3}\cdot n\right)\cdot\frac{r \left(\frac{L^2}{r_{s0}^2}\right)}{qr_{s0}^2\left(1-\frac{r^2}{r_{s0}^2}\right)^2} , \\
\frac{d r}{d \tau}=-\frac{\exp\left(\frac{A}{r}-\frac{B}{r^3}\cdot n\right)}{qr} \cdot \left[1 + \frac{r}{2L}\cdot\frac{L^2}{r_{s0}^2\left(1-\frac{r^2}{r_{s0}^2}\right)} \right]+\frac{\exp\left(-\frac{A}{r_{s0}}-p_0\right)}{qr},\\
\frac{d n}{d \tau}=-\frac{n}{r^2} \cdot \left[1 + \frac{r}{2L}\cdot\frac{L^2}{r_{s0}^2\left(1-\frac{r^2}{r_{s0}^2}\right)} \right]+r\cdot \frac{p_0}{B}  \\
\end{cases}
\end{equation}
\begin{figure}
  \centering
  \includegraphics[width=5.5 cm, height=5.5 cm]{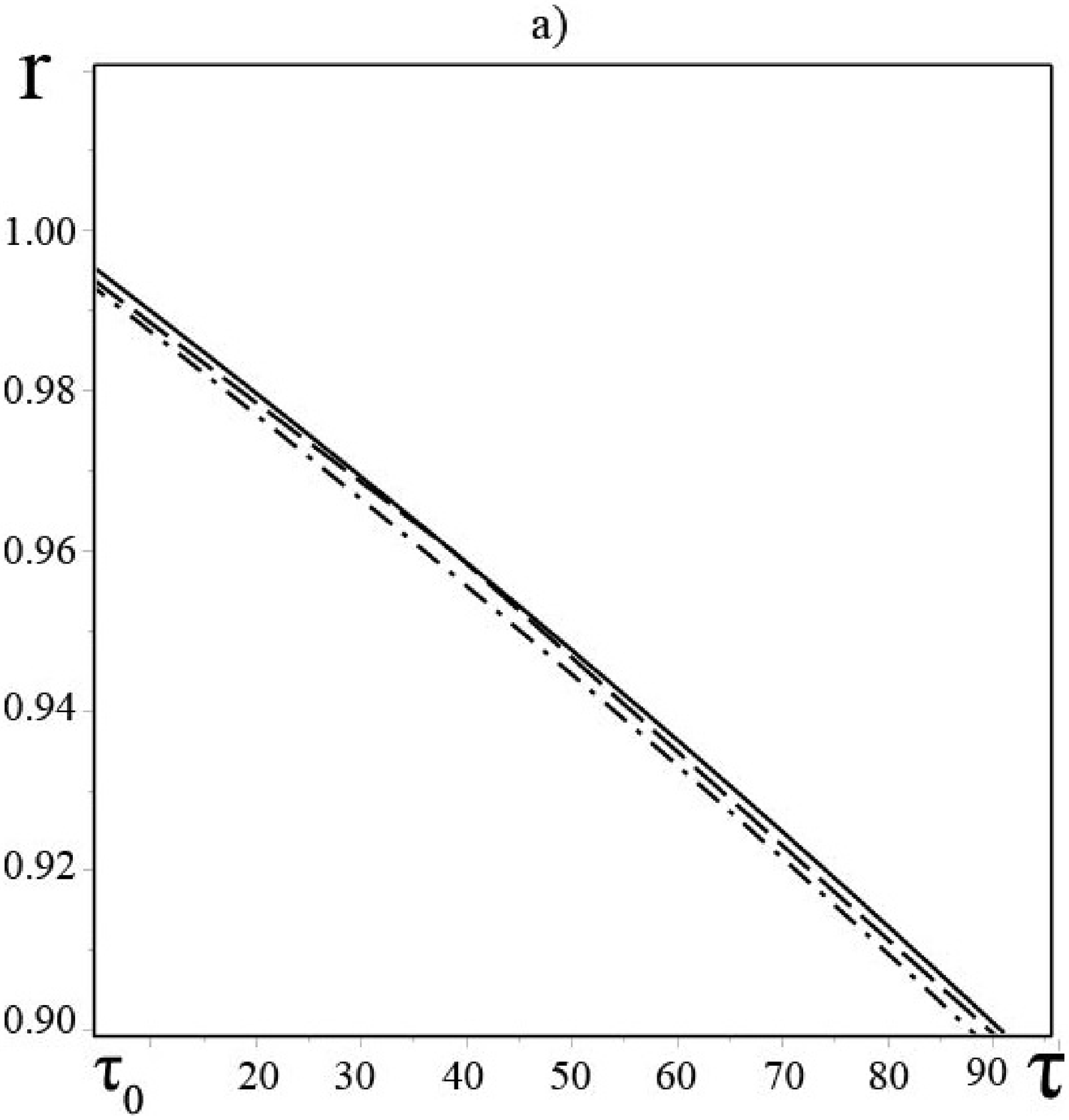}
	\includegraphics[width=5.5 cm, height=5.5 cm]{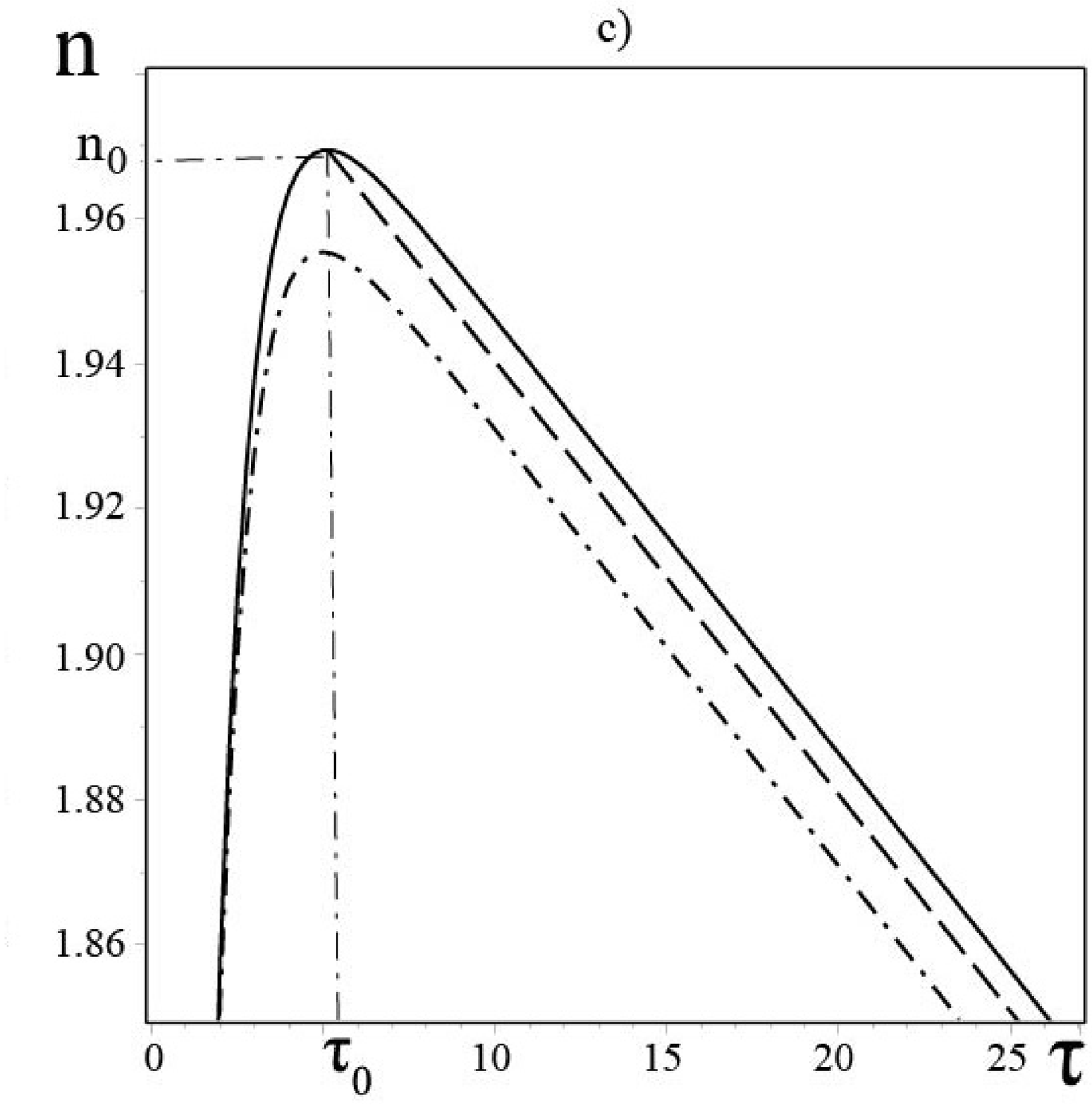}
	\includegraphics[width=5.6 cm, height=5.5 cm]{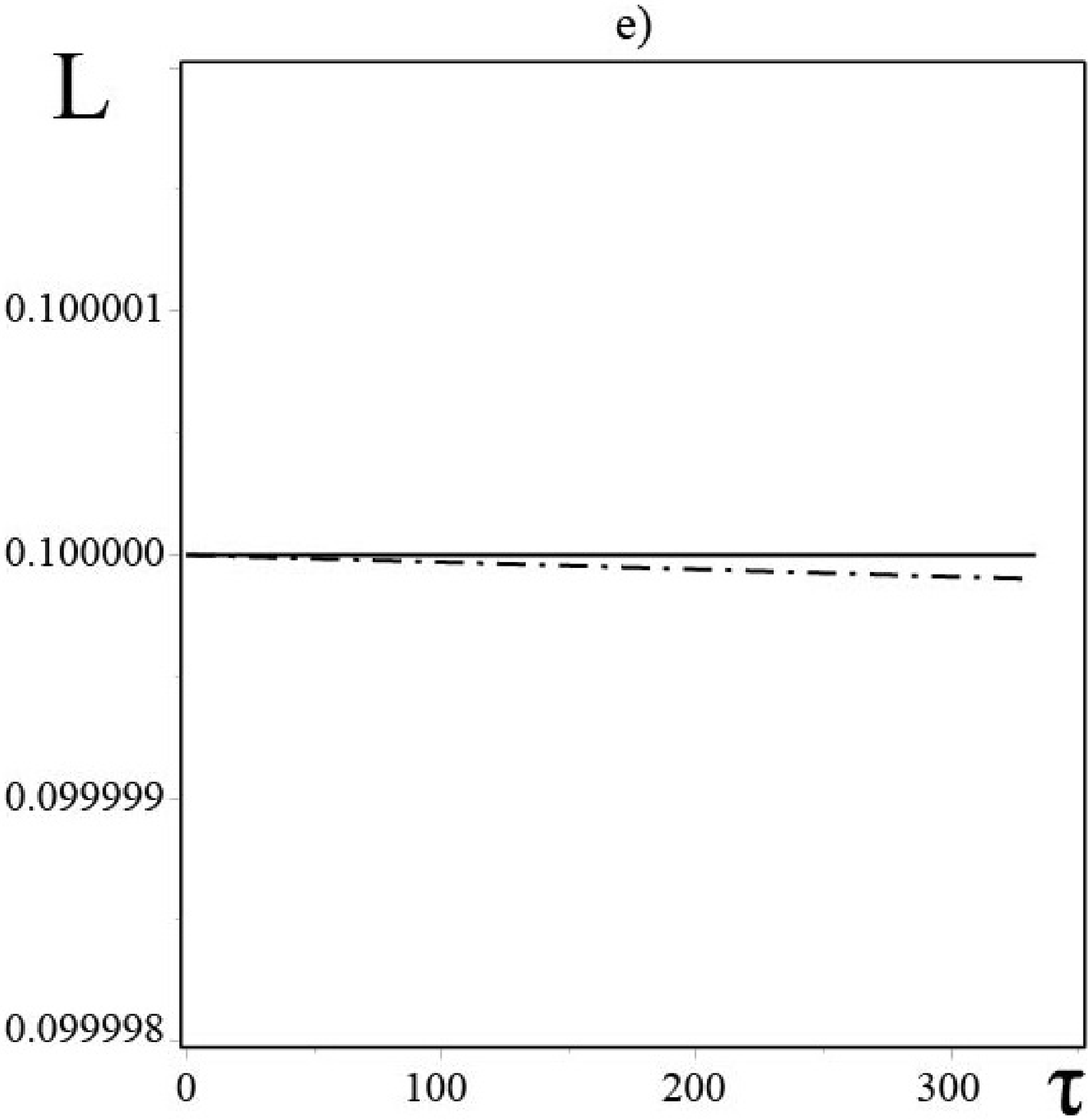}\\
	\includegraphics[width=5.5 cm, height=5.5 cm]{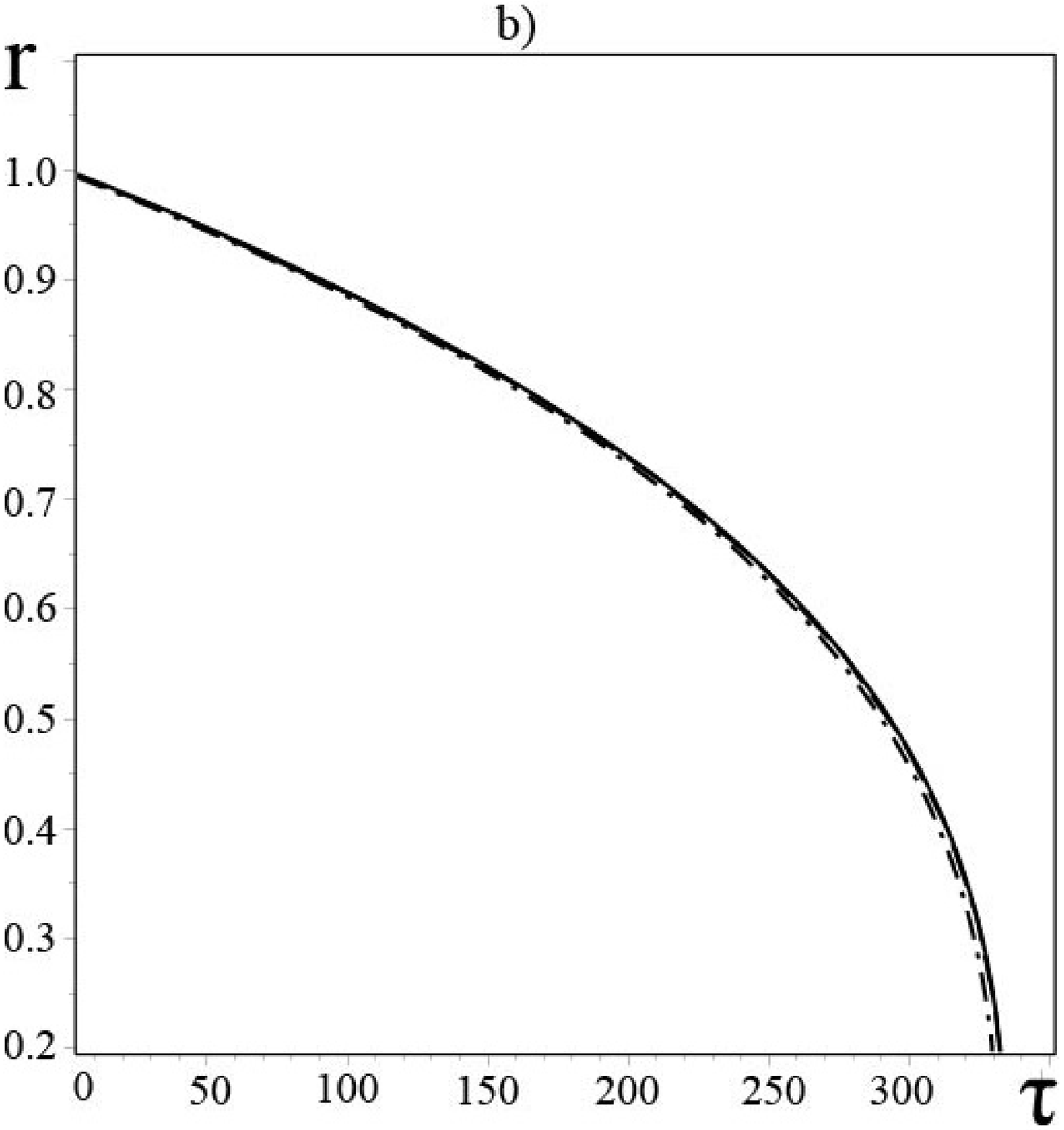}
	\includegraphics[width=5.5 cm, height=5.5 cm]{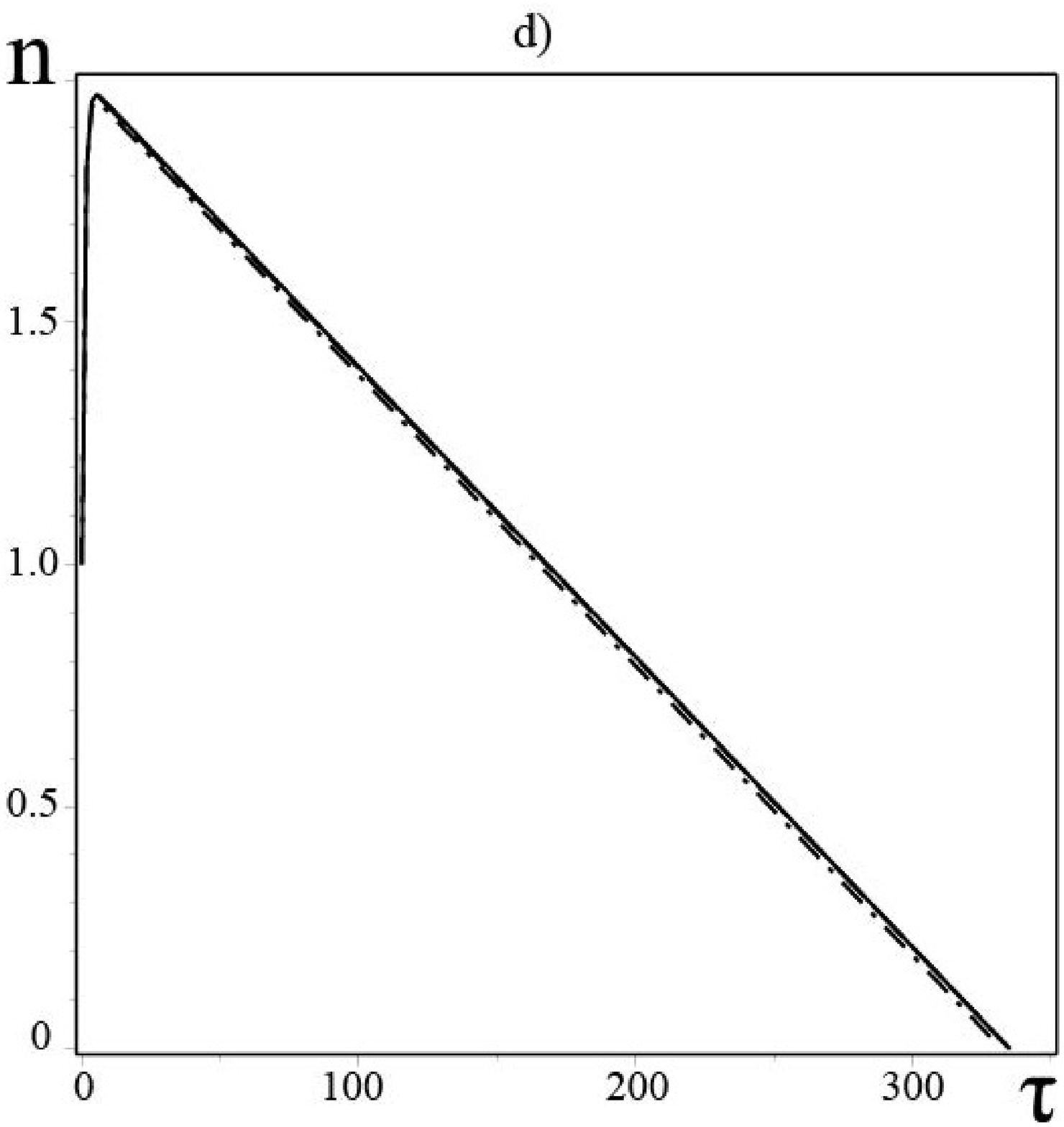}
	\includegraphics[width=5.7 cm, height=5.5 cm]{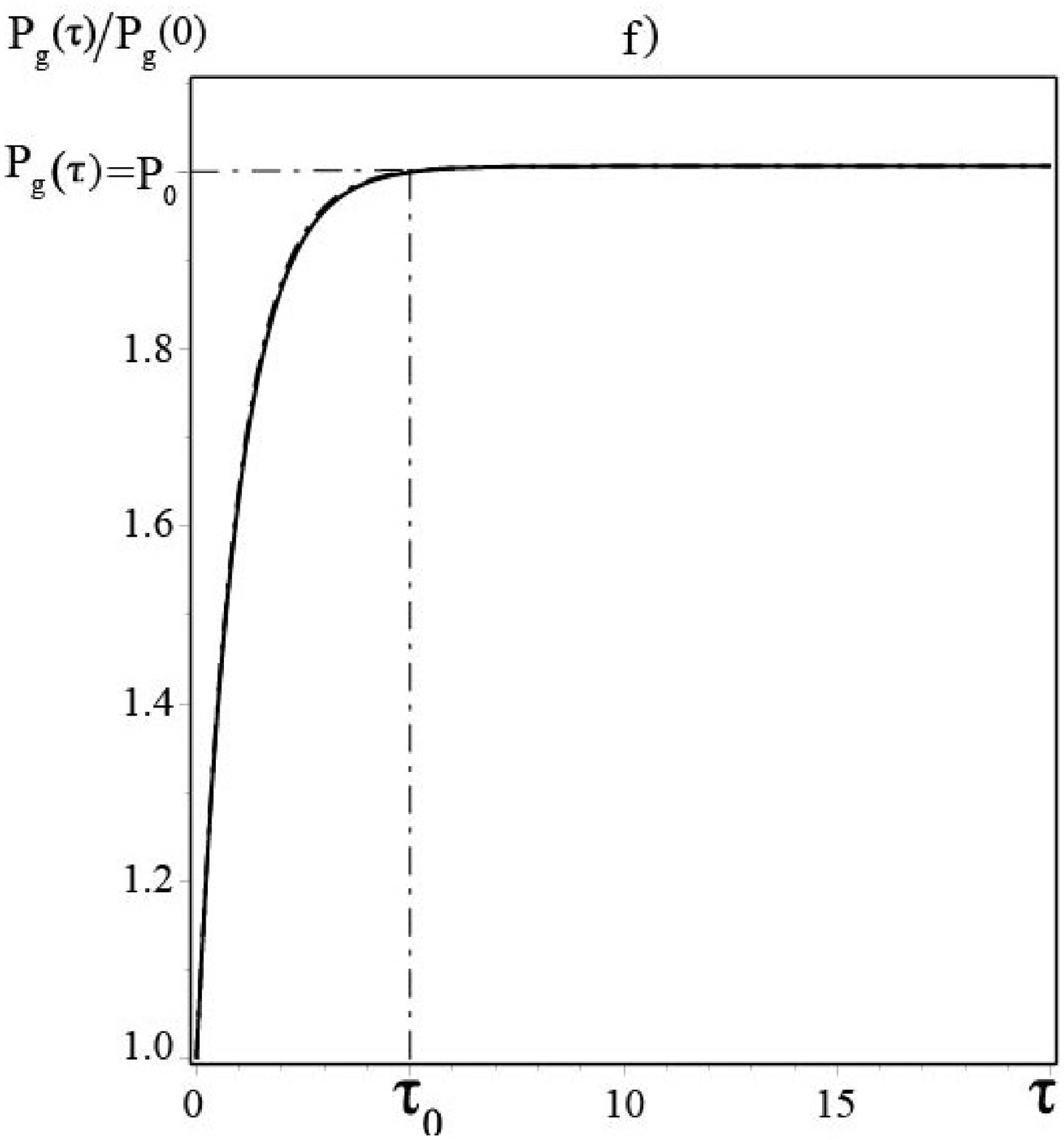}\\
   \caption {a,b) plots of dependence of pore radius $r$ on time $\tau$; c,d) -- the plot of dependence of gas atom number $n$ in the pore $n$ on time $\tau$; e) plots of the dependence of distance $L$ from time $\tau$; f) -- the plot of dependence of gas pressure in the pore $P_g(\tau)/P_g(0)$ on time $\tau$. Dashed-and-dotted lines in the plots corresponds to solutions of complete equation set (\ref{eq5A}). Solid line indicates the numerical solution of approximate  equation set  (\ref{eq11A}), while dashed line indicates analytical solutions of approximate equation set (\ref{eq11Aa}). All solutions are obtained for initial conditions $r|_{\tau=0}=1$, $r_s|_{\tau=0}=100$, $L|_{\tau=0}=0.1$, $n|_{\tau=0}=1$, $A=10^{-3}$, $p_0 = 5\cdot 10^{-3}$, $q=1$, $B=2.5\cdot 10^{-3}$.}
    \label{fg3}
\end{figure}
Numerical solutions demonstrate the behavior similar to the previous case. Gas pressure in the pore reaches the value of the outer pressure and all consequent pore evolution occurs under constant gas pressure in the pore. Thus, it would be natural to investigate in more details regime with stabilized gas pressure in the pore in this case also.

Let us suppose that, after the lapse of time $\tau_0$,  gas  pressure $P_g$ established in the pore equal to the outer pressure  $P_0$. Correspondingly, the regime  established where the number of gas atoms in the pore and pore radius are related via ideal gas equation (\ref{eq9aa}).
In this case, equation set (\ref{eq11A}) takes on a simple form:
\begin{equation}\label{eq11Aa}
\begin{cases}
L(\tau)\approx L|_{\tau=0}=L_0 \\
\frac{d r}{d \tau}=-\frac{A}{qr^2}\cdot\left[1 + \frac{r L_0}{2r_{s0}^2} \right]\cdot e^{-p_0} \\
\frac{d n}{d \tau}=-\frac{n}{r^2} \cdot \left[1 + \frac{r L_0}{2r_{s0}^2} \right]+r\cdot \frac{p_0}{B}  \\
\end{cases}
\end{equation}
This equation set coincides completely with equations (\ref{eq10aa}) and is obtained for the similar conditions
$$  r \ll r_{s0}, \quad A \ll r, \quad A \gg \frac{r^2L_0}{2r_{s0}^2}.  $$
Evidently, analytical solution of equations (\ref{eq11Aa}) will also coincide with the solution of (\ref{eq11aa}). Pore life time in this regime can be found via considerations similar to the previous ones. Here we use solutions of (\ref{eq11aa}) in the form
\[\Delta \tau \approx \frac{q}{A}e^{p_0}\left(\frac{1}{3} - \frac{L_0}{8r_{s0}^2} \right)\]
Here the small correction of the order of $\frac{L_0}{r_{s0}^2}$ is taken into account.

 In Fig. \ref{fg3} there are presented numerical solutions of simplified equation set (\ref{eq11Aa}) (solid line), complete equation set   (\ref{eq5A}) (dashed and dotted line), sa well as analytical solution of equations (\ref{eq11Aa}) (dashed line) for time $\tau \geqslant \tau_0$.   The calculations are performed for initial conditions that satisfy inequality (\ref{eq3A}): $r|_{\tau=0}=1$, $r_s|_{\tau=0}=100$, $L|_{\tau=0}=0.1$, $n|_{t=0}=1$, $A=10^{-3}$, $p_0 = 5\cdot 10^{-3}$, $q=1$, $B=2.5\cdot 10^{-3}$. In Fig. \ref{fg3}($\textrm{a}$-$\textrm{d}$) very good agreement is shown of numerical solutions of equations (\ref{eq11Aa}) and (\ref{eq5A}) with analytical solutions of equations (\ref{eq11Aa}) for pore radius change and the number of gas atoms in the pore. The plots in Fig. \ref{fg3}$\textrm{e}$ demonstrate distance change between pore and granule centers with time for equation set (\ref{eq11Aa}) and equations (\ref{eq5A}) correspondingly. As in the previous case, pore pore displacement is small $L(t) \approx L(0)$. In Fig. \ref{fg3}$\textrm{f}$ evolution of gas pressure  $P_g(\tau)/P_g(0)=n(\tau)/r^3(\tau)$  is shown in the pore during the time $\tau$. It can be seen, that during the time $\tau_0$ gas pressure $P_g(\tau_0)$ is established equal to outer pressure $P_0$, that is $P_g(\tau_0)=P_0=p_0/B=2$. During the entire time of pore healing, gas pressure in the pore is sustained on the level $P_0$: $P_g(\tau \geqslant \tau_0)=P_0$.

\subsubsection{Pore near granule boundary}

Let us now turn to discussing the mode (\ref{eq4A}) when pore is situated close to  granule boundary . In this case, the relation $l/R_s$ is close to 1
\[\frac{l}{R_s}=1-\varepsilon,\]
here $\varepsilon$ is small parameter of asymptotic expansion. Parameter $\varepsilon$ is limited by geometrical inequality (the pore inside the granule)
\[\frac{R}{R_s} \leq \varepsilon .\]
In asymptotic expansion over $\alpha$, let us restrict ourselves by the terms of the order of $\varepsilon$. With account of this notice, parameter $\alpha$ and, correspondingly, bispherical coordinates $\eta_{1,2}$, obtained in the approximation of small pore $R\ll R_s$ and $R \ll l$, take on the form:
\begin{equation}\label{eq12A}
 \alpha \approx \frac{r_{s0}^2}{L}\varepsilon=\frac{r_{s0}^2}{L} \left(1-\frac{L}{r_{s0}}\right), \; \sinh\eta_1=\frac{r_{s0}^2}{rL} \left(1-\frac{L}{r_{s0}}\right),\;\sinh \eta_2=\frac{r_{s0}}{L} \left(1-\frac{L}{r_{s0}}\right). \end{equation}
It can be seen from here, that $\eta_1 \gg \eta_2$ , consequently we can use previous estimates for series sums using formulas (\ref{eq8A}). Substitution of (\ref{eq8A}) and (\ref{eq12A}) into right part of equations  (\ref{eq5A}) gives pore evolution equation under conditions  (\ref{eq9aa}):
\begin{equation}\label{eq13A}
\begin{cases}
L(\tau)\approx L|_{\tau=0}=L_0 \\
\frac{d r}{d \tau}=-\frac{A}{qr^2}\cdot\left[1 + \frac{r L_0}{4r_{s0}(r_{s0}-L_0)} \right]\cdot e^{-p_0} \\
\frac{d n}{d \tau}=-\frac{n}{r^2} \cdot \left[1 + \frac{r L_0}{4r_{s0}(r_{s0}-L_0)} \right]+r\cdot \frac{p_0}{B}  \\
\end{cases}
\end{equation}
\begin{figure}
  \centering
  \includegraphics[width=5.5 cm, height=5.5 cm]{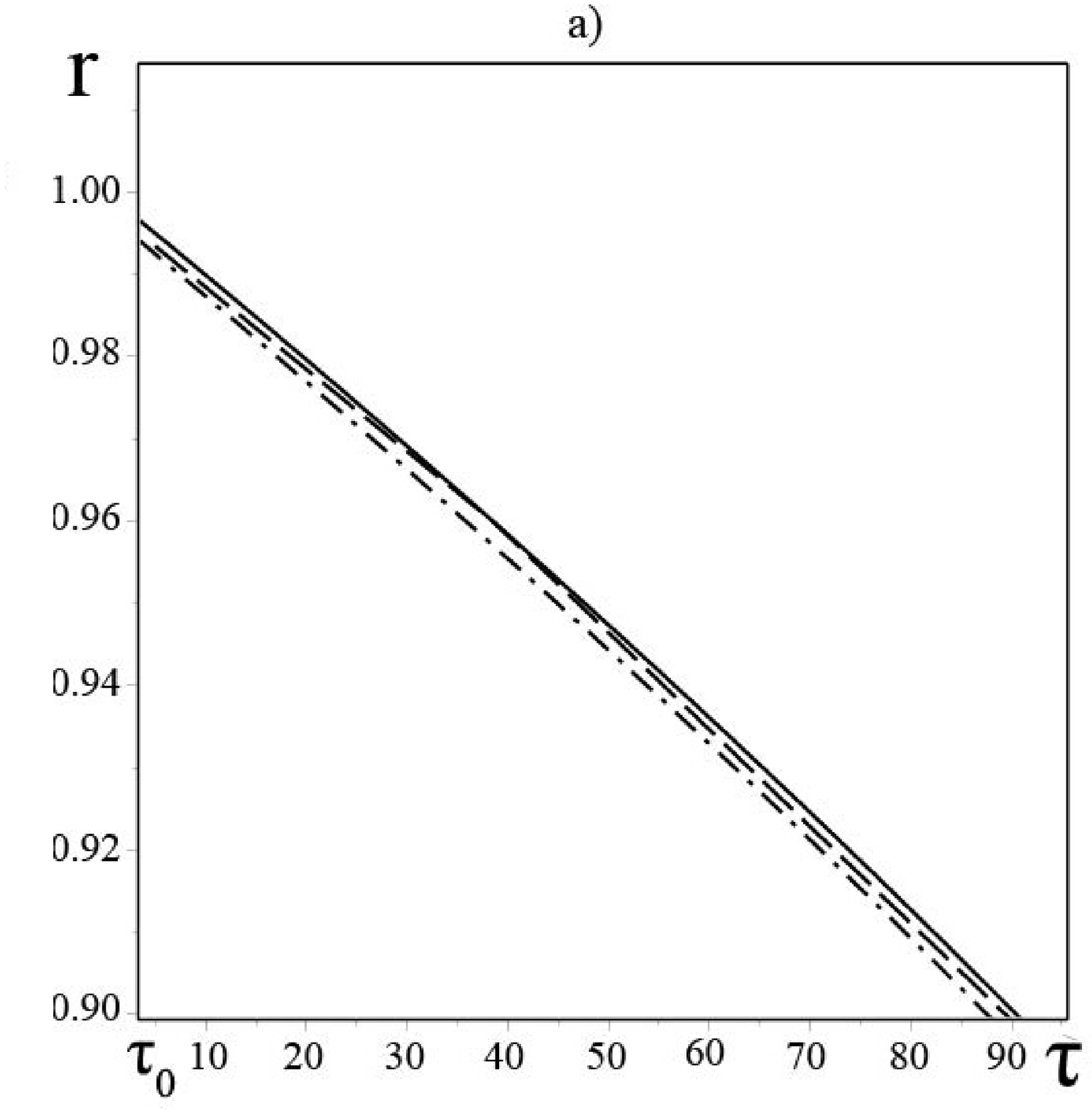}
	\includegraphics[width=5.5 cm, height=5.5 cm]{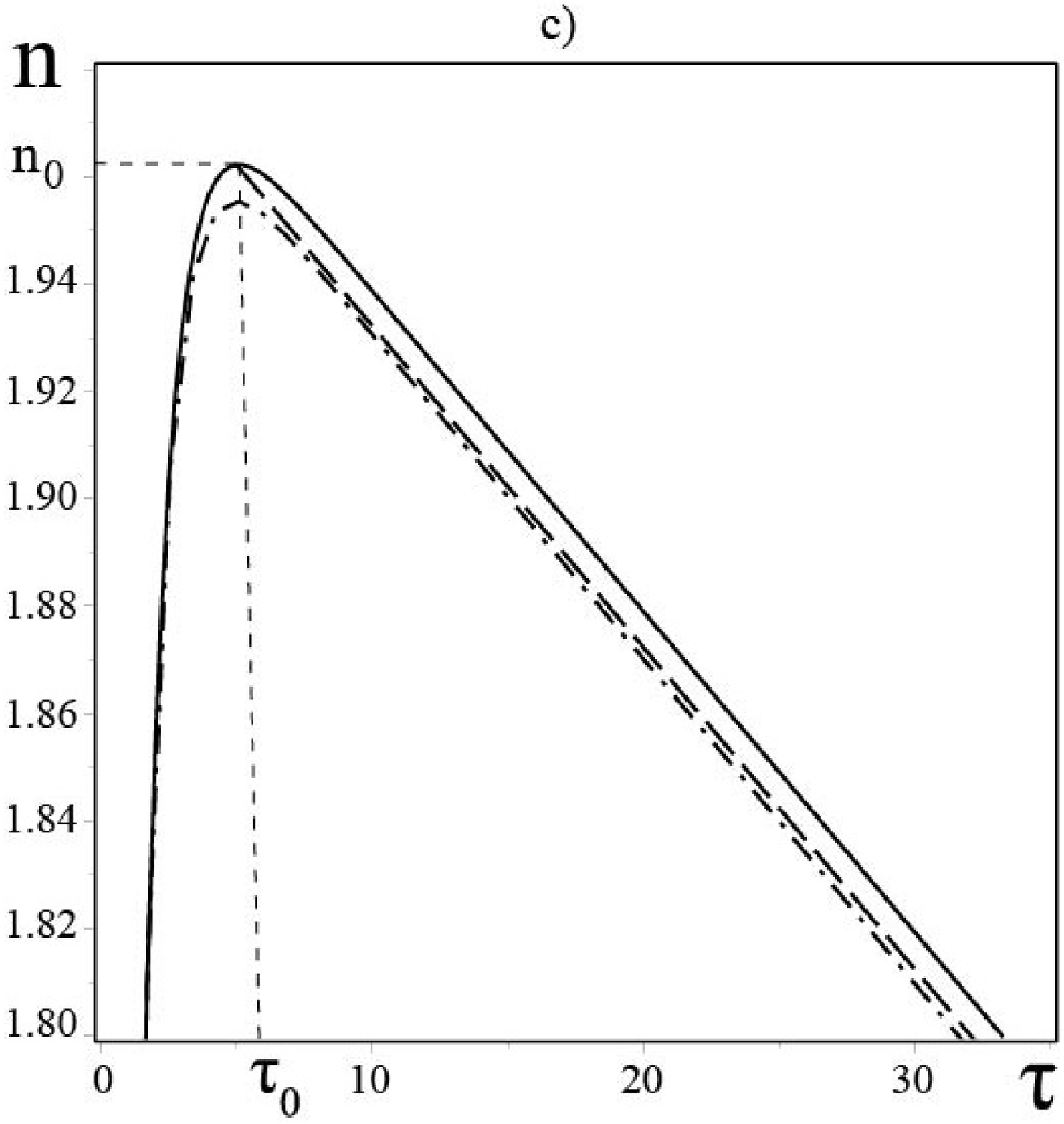}
	\includegraphics[width=5.5 cm, height=5.5 cm]{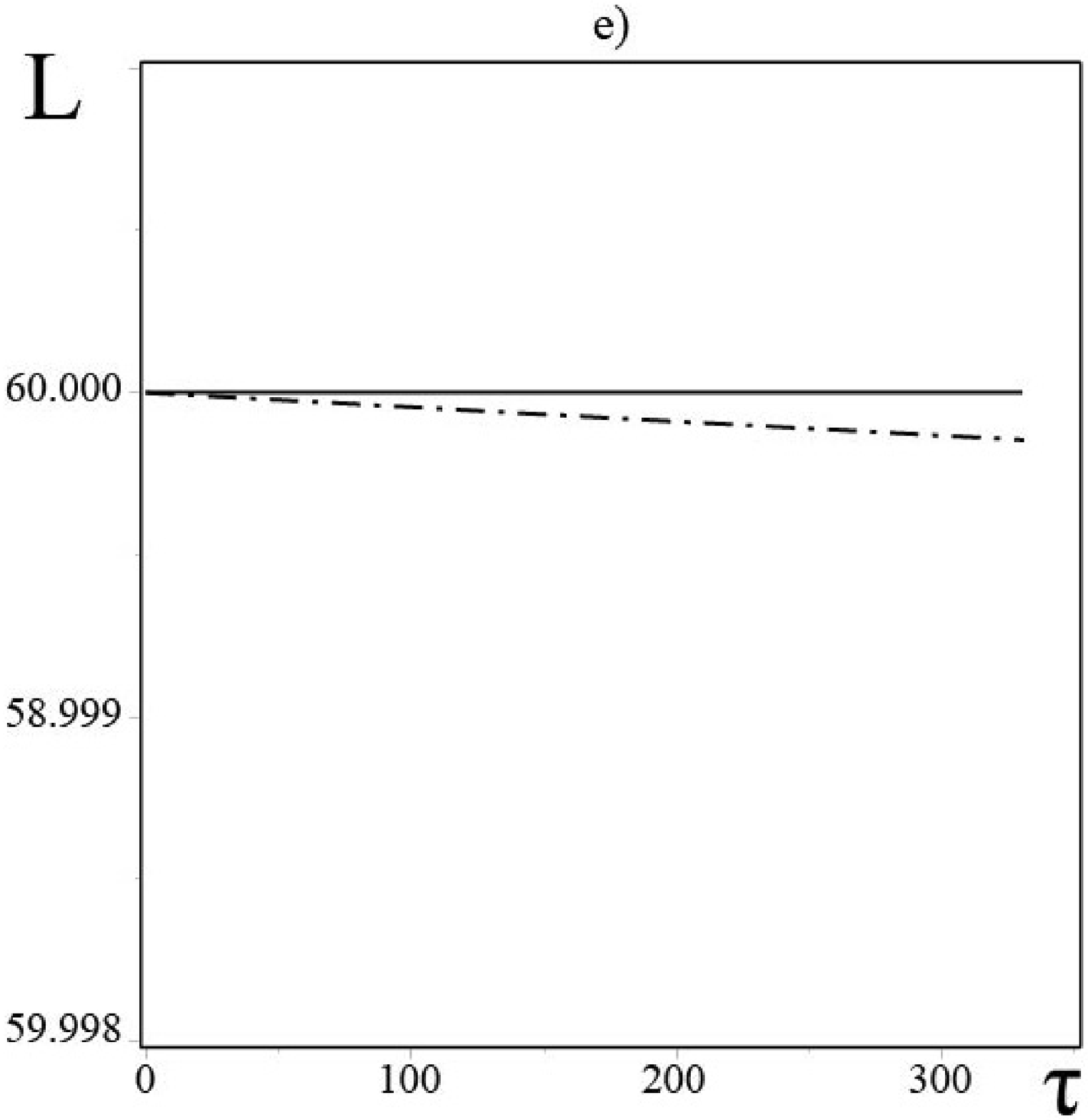}\\
	\includegraphics[width=5.5 cm, height=5.5 cm]{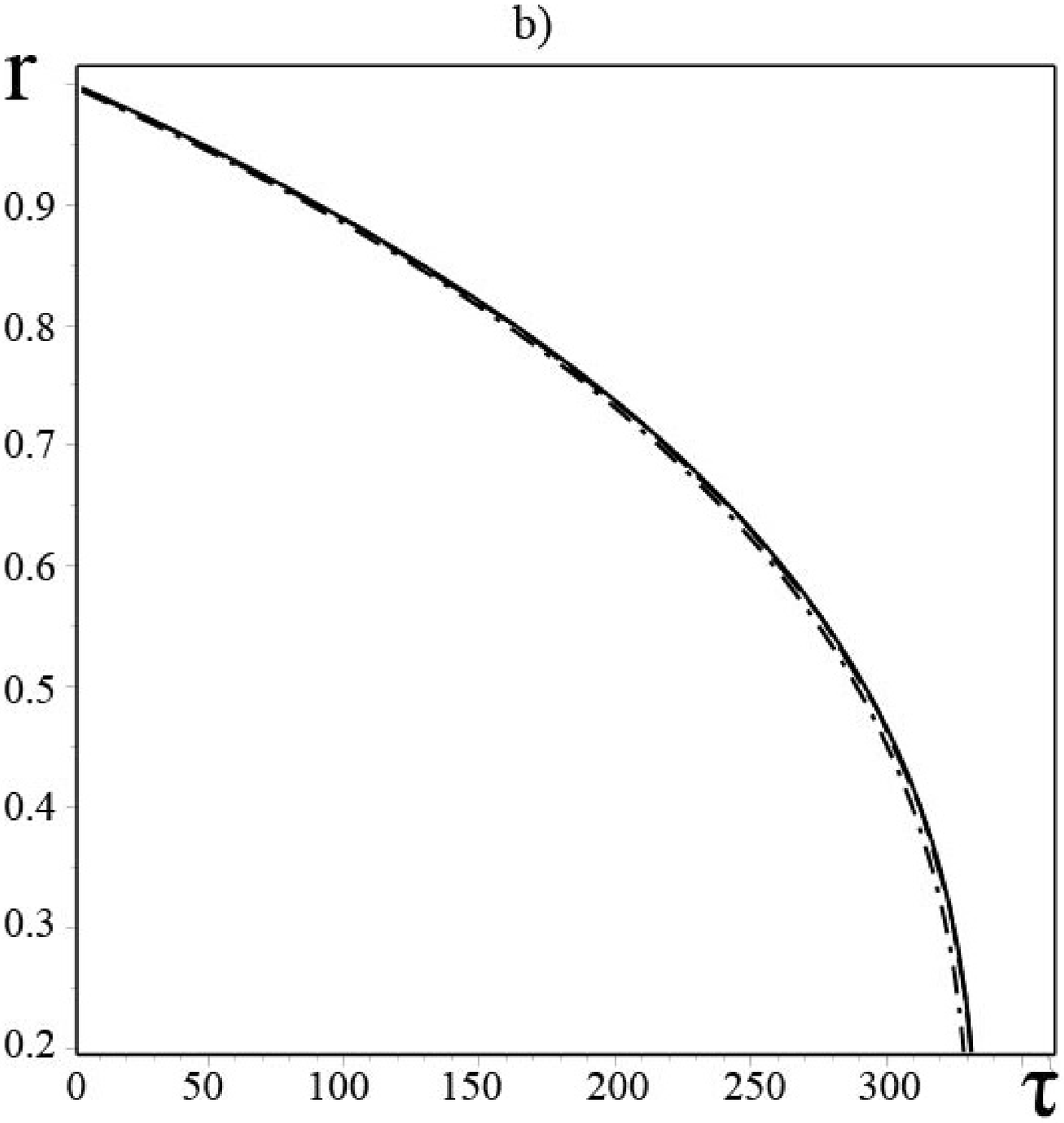}
	\includegraphics[width=5.5 cm, height=5.5 cm]{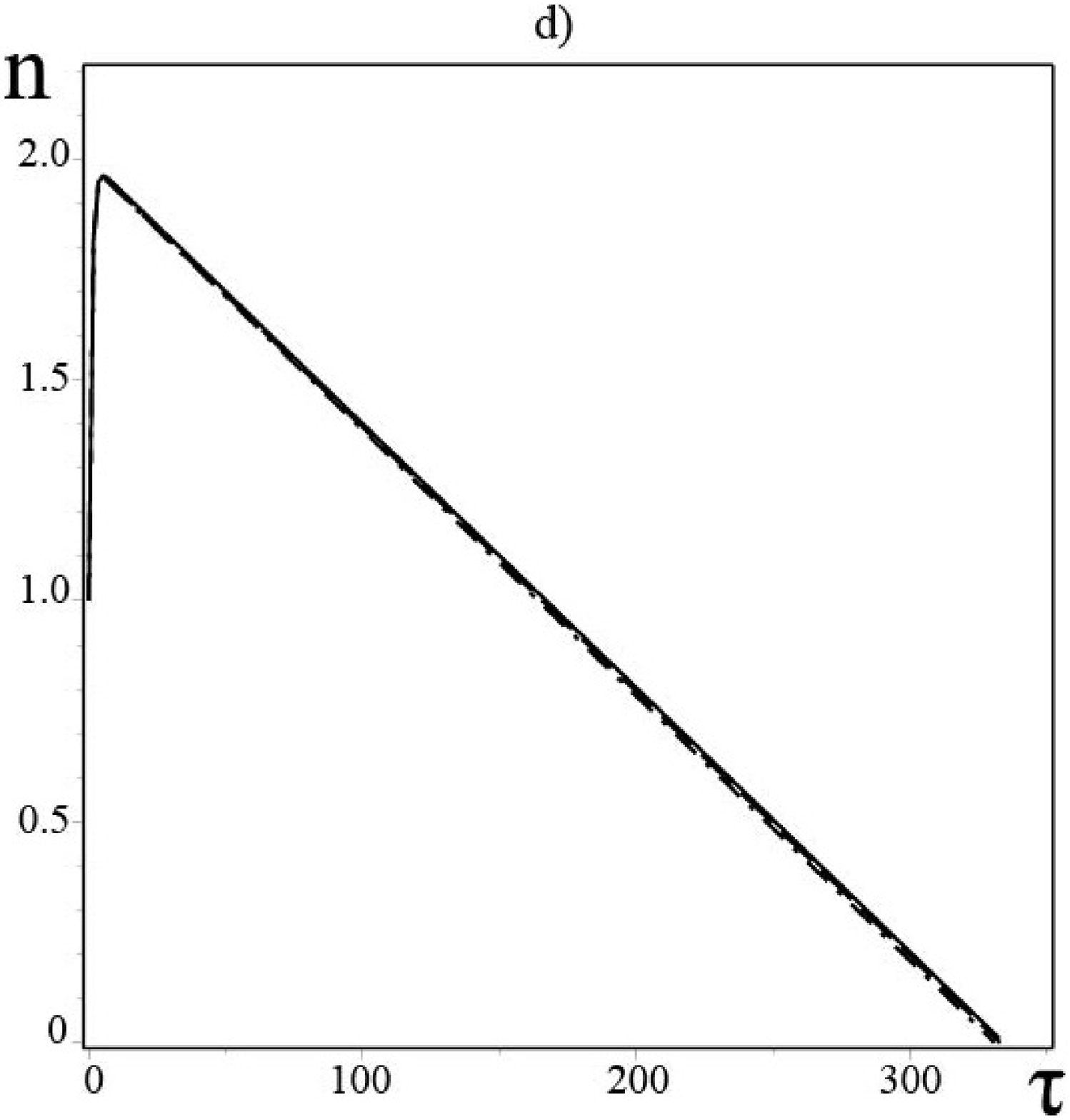}
	\includegraphics[width=5.6 cm, height=5.5 cm]{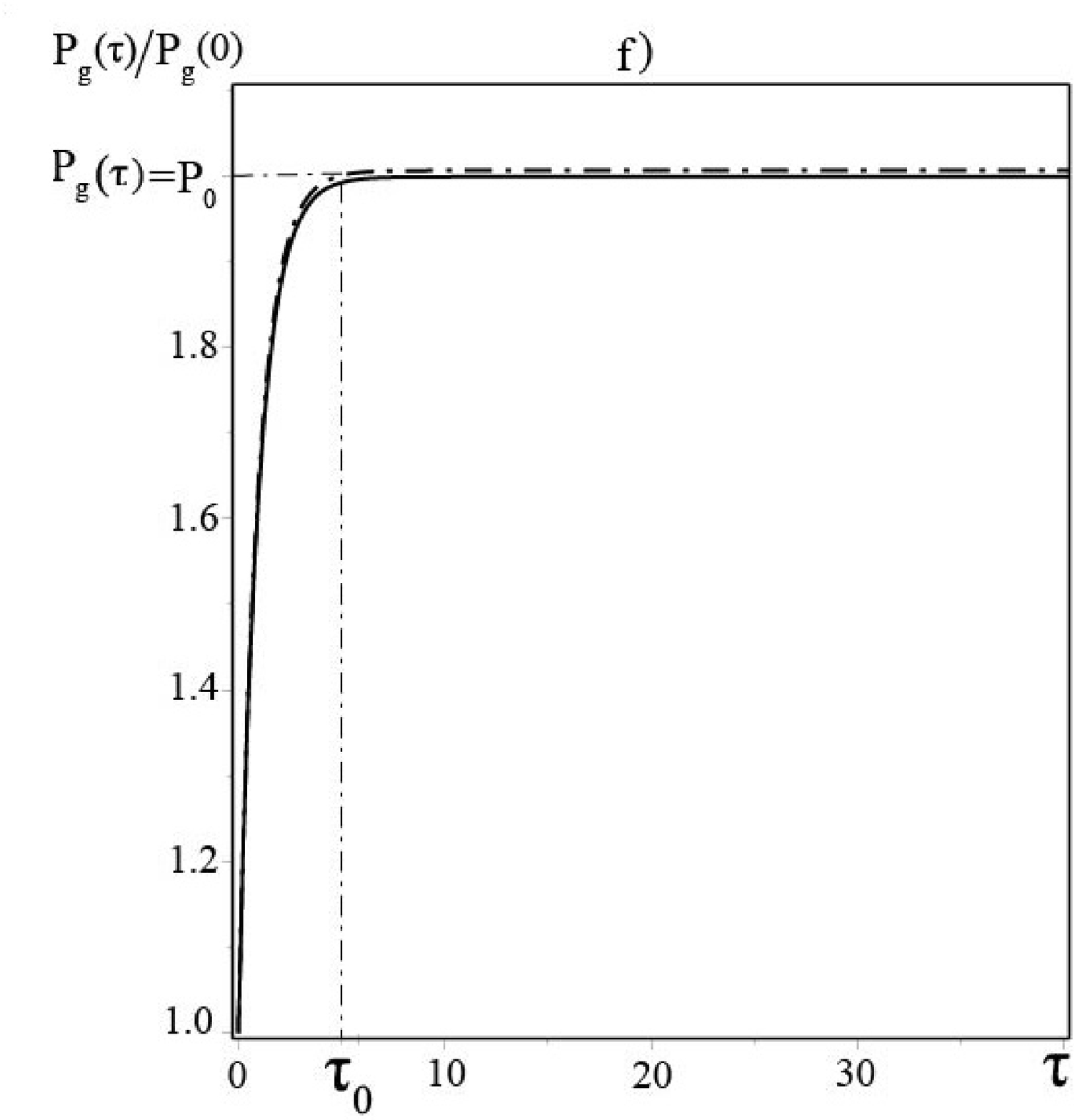}\\
	  \caption {a,b) plots of dependence of pore radius $r$ on time $\tau$; c,d) -- the plot of dependence of gas atom number $n$ in the pore $n$ on time $\tau$; e) plots of the dependence of distance $L$ from time $\tau$; f) -- the plot of dependence of gas pressure in the pore $P_g(\tau)/P_g(0)$ on time $\tau$. Dashed-and-dotted lines in the plots corresponds to solutions of complete equation set (\ref{eq5A}). Solid line indicates the numerical solution of approximate  equation set  (\ref{eq13A}), while dashed line indicates analytical solutions of approximate equation set (\ref{eq13A}). All solutions are obtained for initial conditions $r|_{\tau=0}=1$, $r_s|_{\tau=0}=100$, $L|_{\tau=0}=60$, $n|_{\tau=0}=1$, $A=10^{-3}$, $p_0 = 5\cdot 10^{-3}$, $q=1$, $B=2.5\cdot 10^{-3}$.}
\label{fg4}
\end{figure}
This equation set  is similar to the equations (\ref{eq10aa}) and is obtained for the following conditions:
$$  r \ll r_{s0}, \quad A \ll r, \quad A \gg \frac{r^2L_0}{4r_{s0}(r_{s0}-L_0)} .  $$
Analytical solution for pore radius change  from set (\ref{eq13A}) will coincide with solution (\ref{eq11aa}) if coefficient $\widetilde{\alpha}=L_0/2r_{s0}^2$ is substituted by   $\widetilde \alpha=\frac{L_0}{4r_{s0}(r_{s0}-L_0)}$. Thus, the characteristic pore  healing time is
\[\Delta \tau \approx \frac{q}{A}e^{p_0}\left(\frac{1}{3} - \frac{L_0}{16 \, r_{s0}(r_{s0}-L_0)} \right)\]
The difference with the previous case depends on the small correction value.

In Fig. \ref{fg4} are shown numerical solutions of complete (\ref{eq5A}) (dashed-and-dotted line)  and approximate (\ref{eq13A}) (solid line) equation set , as well as analytical solution of equations (\ref{eq13A}) (dashed line) for time interval $\tau \geqslant \tau_0$.  The calculations were conducted for the same initial conditions $r|_{t=0}=1$, $r_s|_{t=0}=100$, $L|_{t=0}=60$, $n|_{t=0}=1$, $A=10^{-3}$, $p_0 = 5\cdot 10^{-3}$, $q=1$, $B=2.5\cdot 10^{-3}$. It can be seen in Fig. \ref{fg4}($\textrm{a}$-$\textrm{c}$), that numerical solutions of equations (\ref{eq13A})  coincide with analytical solution (\ref{eq13A}) for the time dependences of pore  radius as well as of  gas atoms number in the pore. Numerical solutions of approximate equation set  (\ref{eq13A}) well agrees with numerical solutions of the exact equation set  (\ref{eq5A}).  In Fig. \ref{fg4}$\textrm{e}$ the plots are shown of time change of the distance between  pore and granule centers. It can be seen, that  pore displacement towards granule center is neglected in approximate equation set  (\ref{eq13A}), that is $L(\tau) \approx L_0$. However, for the complete equation set (\ref{eq5A}), the displacement $L(\tau)$  during pore healing time is quite small. In Fig. \ref{fg4}$\textrm{f}$, the dependence is shown of gas pressure  $P_g(\tau)/P_g(0)=n(\tau)/r^3(\tau)$ in the pore on time $\tau$. It can be seen in the figure, that during time $\tau_0$  the  gas pressure $P_g(\tau_0)$ is established equal to outer pressure $P_0$, that is $P_g(\tau_0)=P_0=p_0/B=2$. Thus, during the whole time time of pore healing, gas pressure in it is sustained on the level  $P_0$: $P_g(\tau \geqslant \tau_0)=P_0$.

Thus, it is characteristic for small pores that pore volume decreases linearly with characteristic rate, that is weakly dependent on pore relative position. Gas pressure in the pore increases up to the value of the outer one and is sustained on this level during the whole time of pore healing. Starting from the moment when constant pressure is established, the number of gas atoms in the pore also decreases linearly with time. The life time of small pores is defined by universal relation (\ref{14bb}).

\subsection{Large pores}

Let us now proceed to discussing the evolution of large pores. Let us begin with the notion, that asymptotic mode 
\begin{equation}\label{eq1B}
R/R_s \cong 1,\quad l/R_s \cong 1,\quad  R/l \cong 1. \end{equation}
is not, in fact, realized. Indeed, let us take into account the closeness of the two firs relations to the unity 
\begin{equation}\label{eq2B}
\frac{R}{R_s}=1-\varepsilon_1,\quad \frac{l}{R_s}=1-\varepsilon_2, \end{equation}
where $\varepsilon_1 \ll 1$ and $\varepsilon_2 \ll 1$ are small parameters. Substituting  (\ref{eq2B}) into geometrical condition (\ref{eq1A}), we find  $1 \leq \varepsilon_1+\varepsilon_2$. Since $\varepsilon_{1,2}$ are small parameters, this inequality does not hold. Thus, mode (\ref{eq1B}) is not compatible with geometrical condition (\ref{eq1A}).

Let us consider the valid regime of large pore evolution when relations between values  $R$, $R_s$, $l$ are the following:
\begin{equation}\label{eq3B}
  \quad R/R_s \cong 1,\quad R \gg l,\quad  R_s \gg l. \end{equation}  
Taking into account general character of evolution of gas-filled pore, one should note that this mode for large pore  is realized for some initial time and, ultimately, changes for small pore mode. In other words, large pore mode is of transitional character towards small pore mode. Therefore, in the further consideration we use only the condition $ R_s \gg l$, that is satisfied for all pore healing times.

By virtue of the volume conservation low for granule material, the relation (\ref{eq26}) is performed. In dimensionless variables it has the form
\begin{equation}\label{eq4B}
r_s(\tau)=\left(r_s^3(0)-r^3(0)+r^3(\tau) \right)^{1/3}=\left(V+r^3(\tau) \right)^{1/3} \end{equation}
\begin{figure}
  \centering
  \includegraphics[width=5.5 cm, height=5.5 cm]{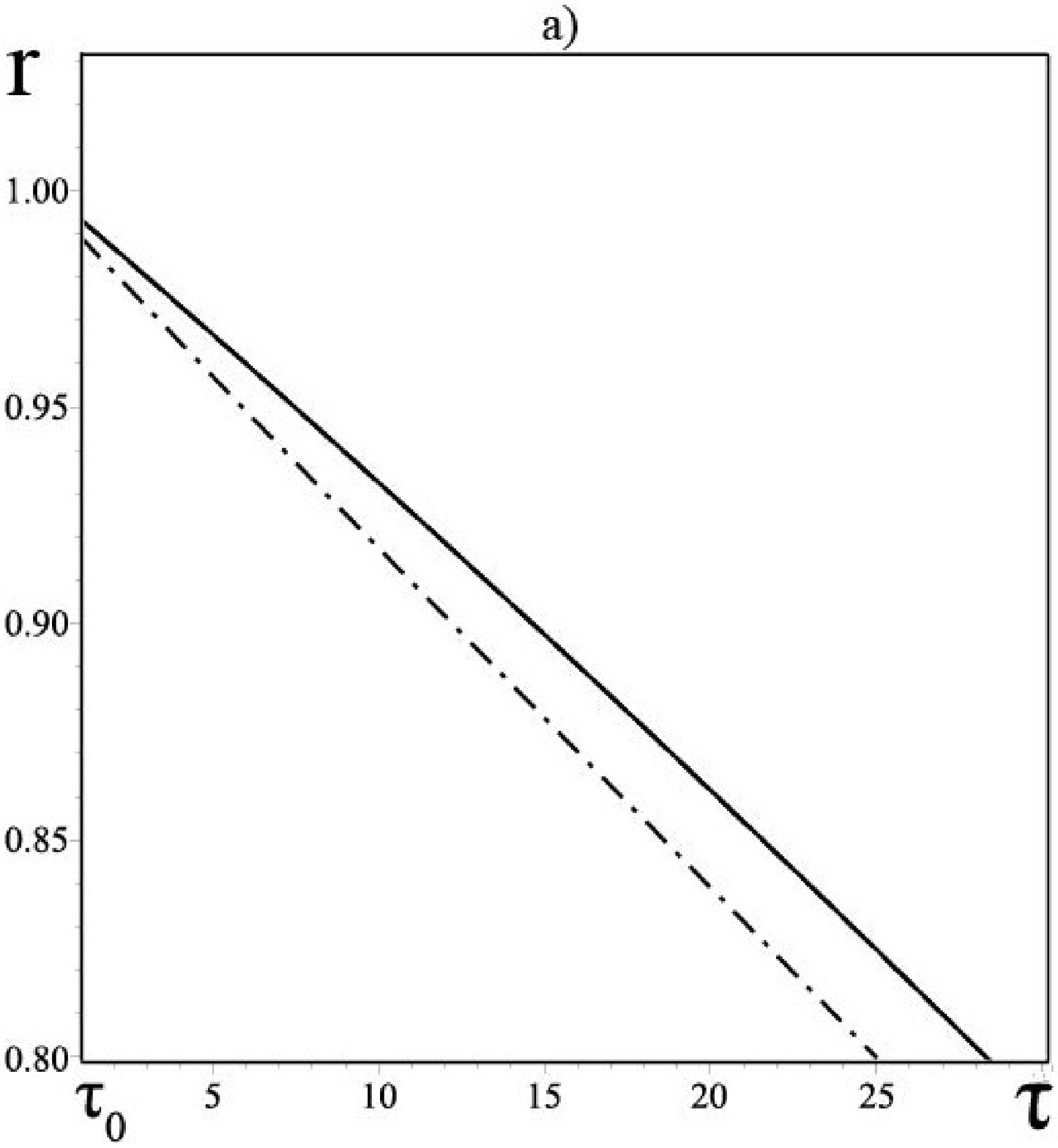}
	\includegraphics[width=5.5 cm, height=5.5 cm]{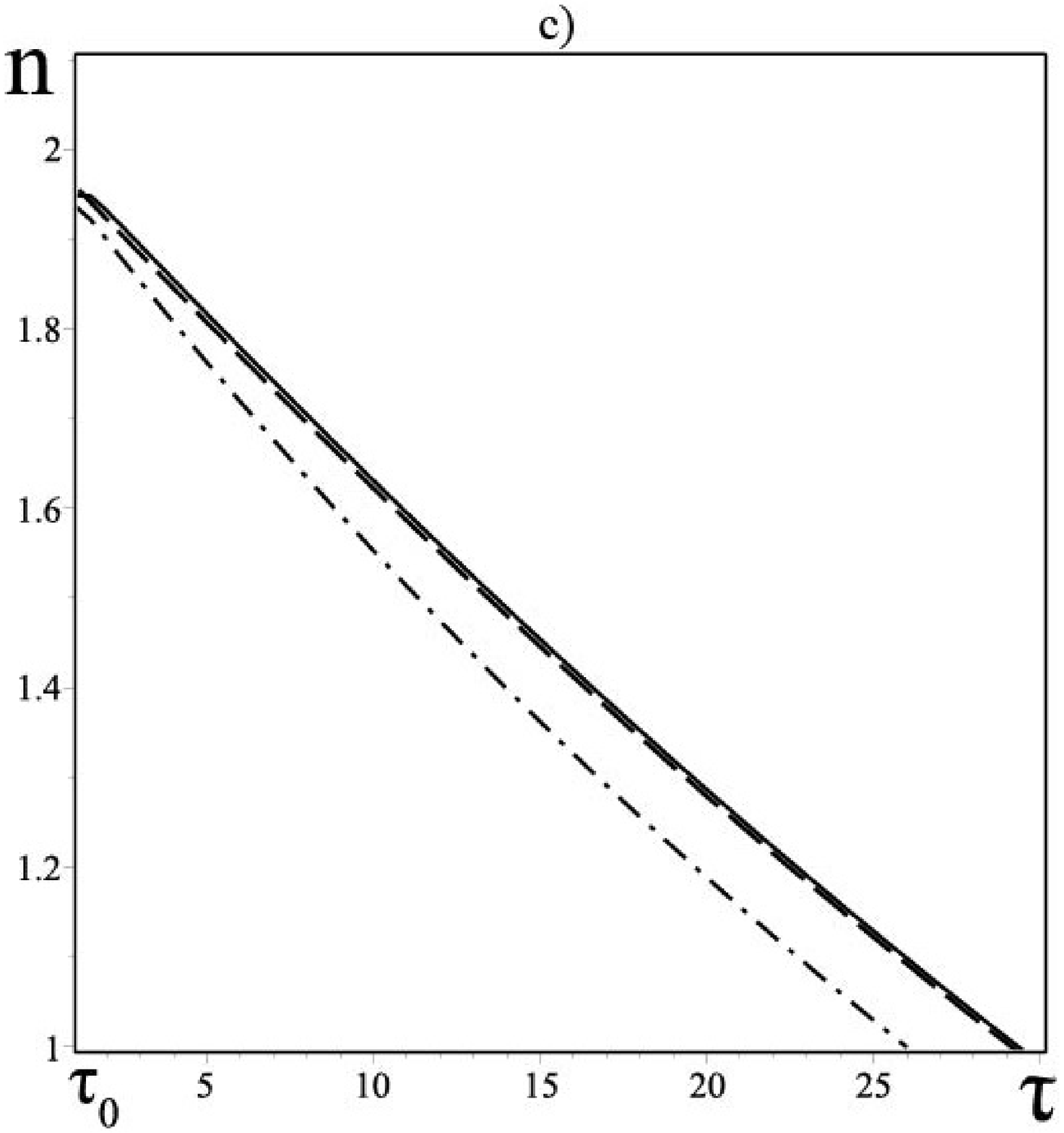}
	\includegraphics[width=5.5 cm, height=5.5 cm]{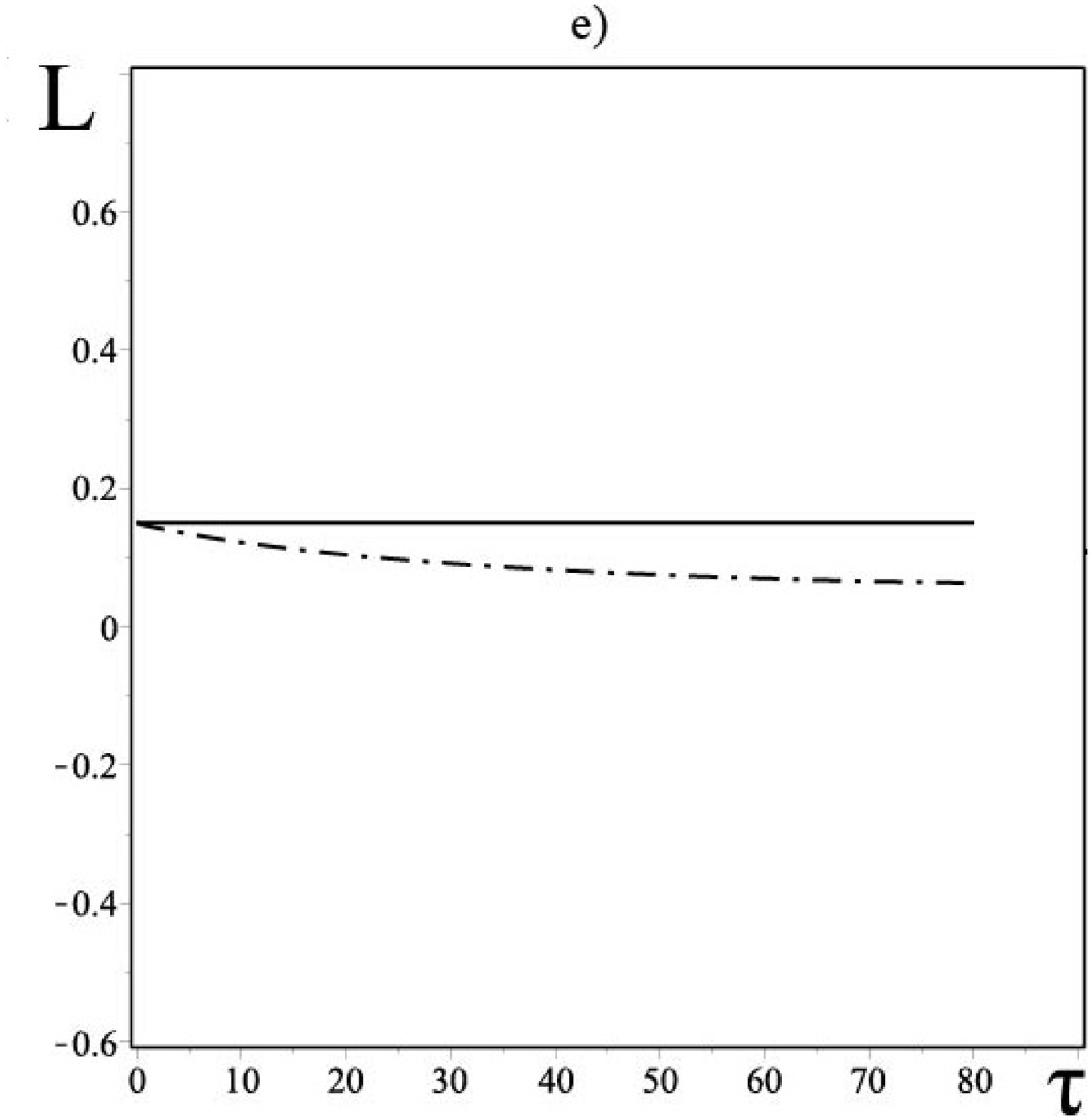}\\
	\includegraphics[width=5.3 cm, height=5.5 cm]{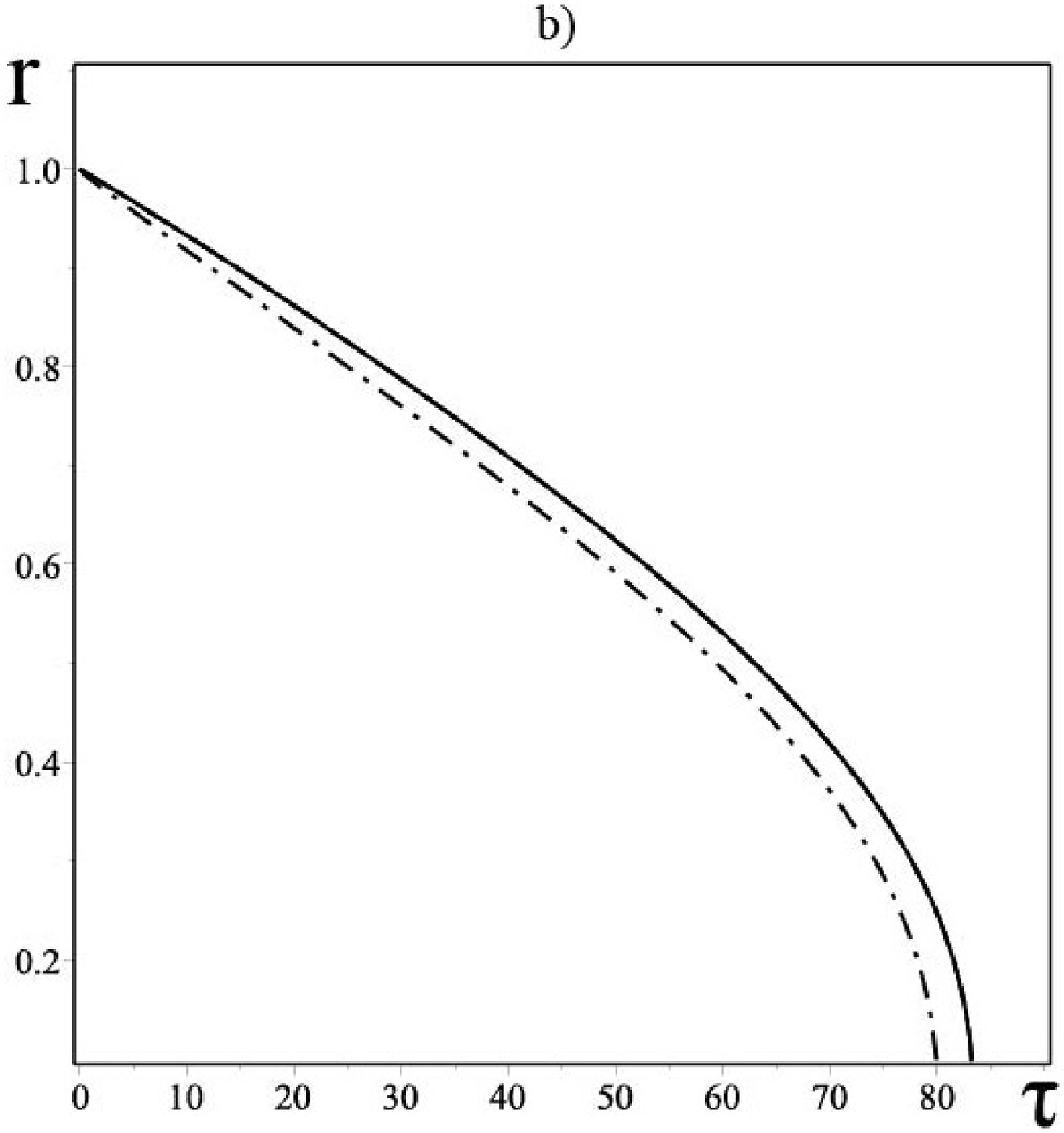}
	\includegraphics[width=5.5 cm, height=5.5 cm]{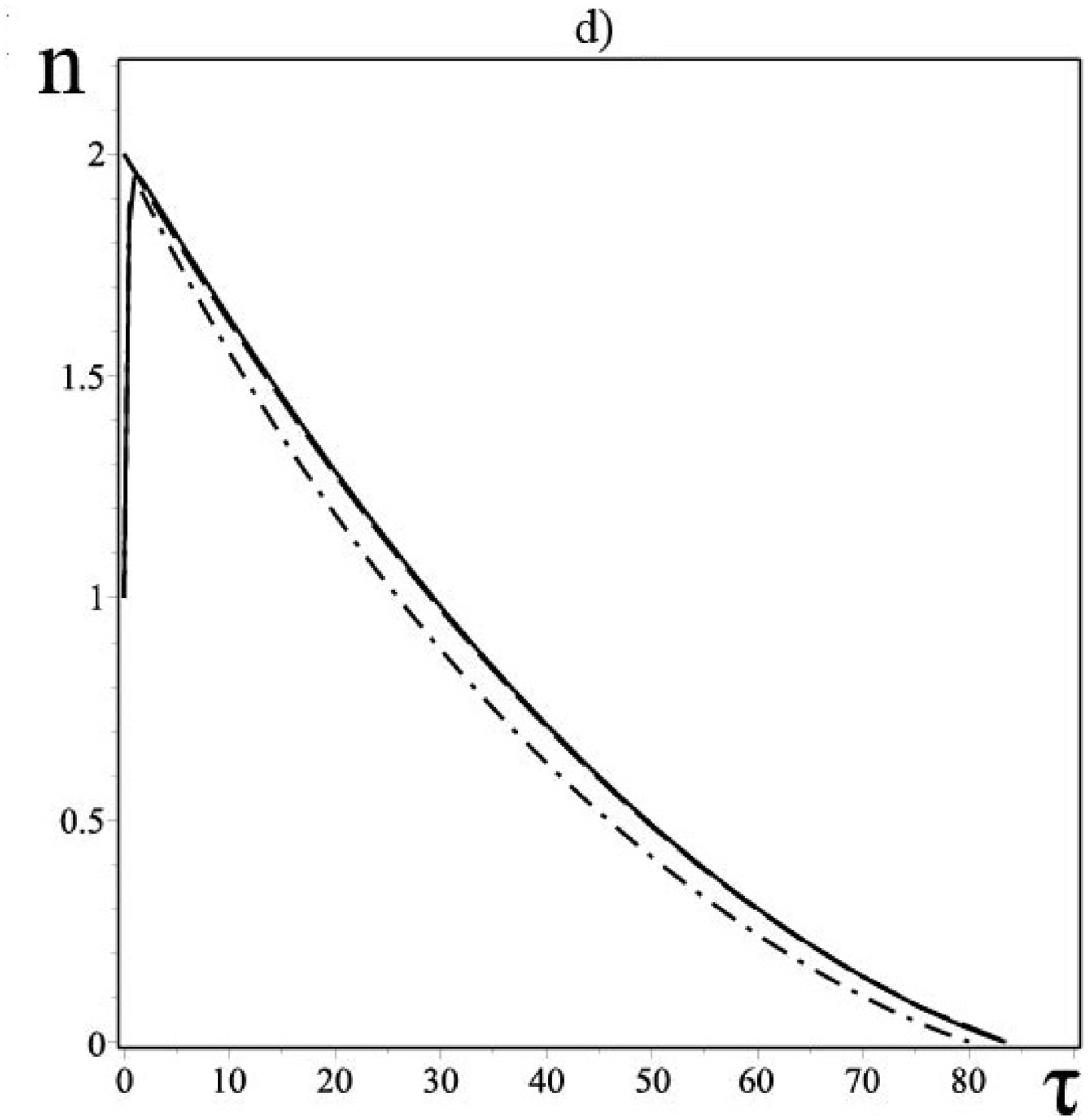}
	\includegraphics[width=5.5 cm, height=5.5 cm]{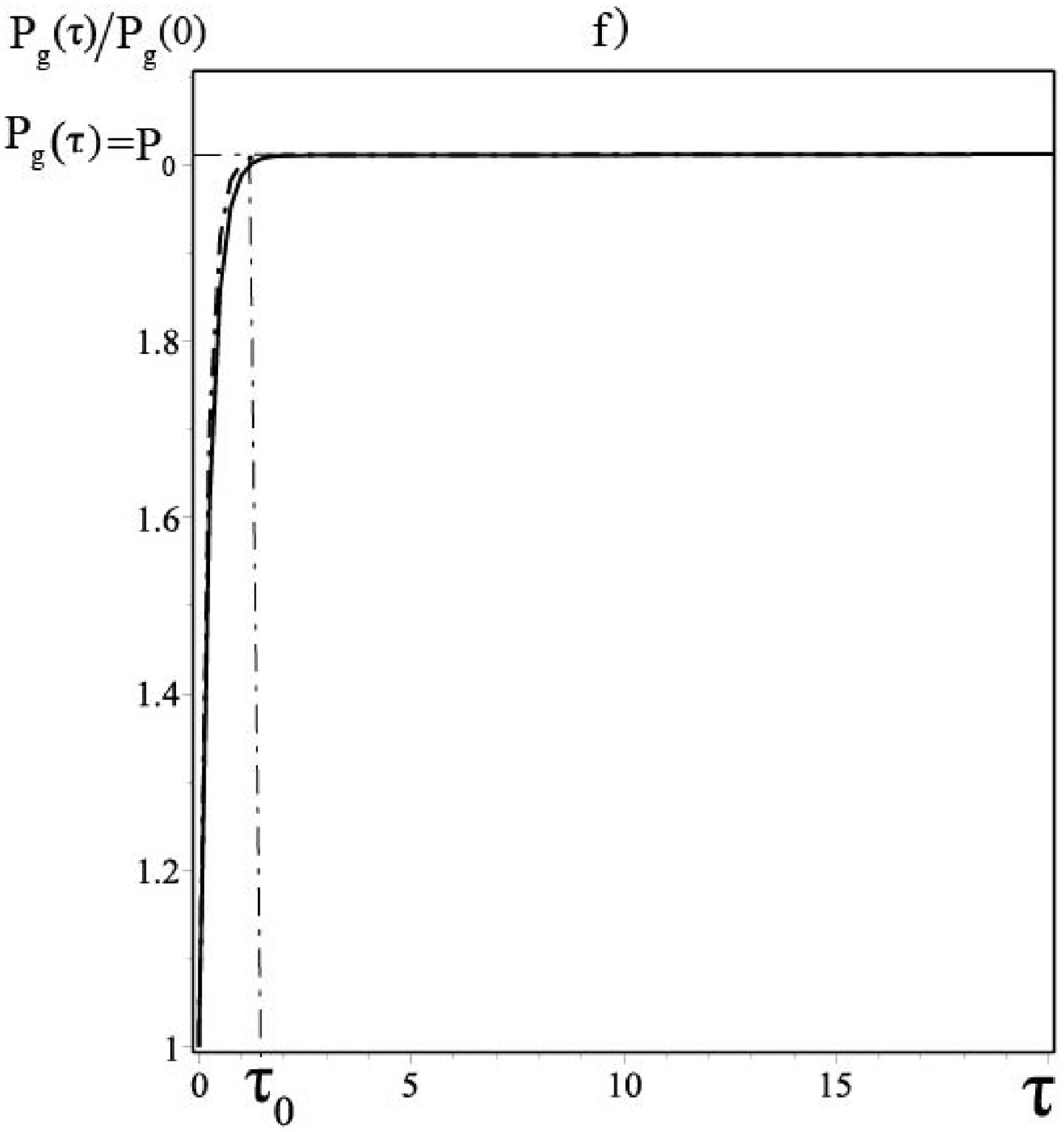}\\
  \caption {a,b) plots of dependence of pore radius $r$ on time $\tau$; c,d) -- the plot of dependence of gas atom number $n$ in the pore $n$ on time $\tau$; e) plots of the dependence of distance $L$ from time $\tau$; f) -- the plot of dependence of gas pressure in the pore $P_g(\tau)/P_g(0)$ on time $\tau$. Dashed-and-dotted lines in the plots corresponds to solutions of complete equation set (\ref{eq29}). Solid line indicates the numerical solution of approximate  equation set  (\ref{eq8B}), while dashed line indicates analytical solutions of approximate equation set (\ref{eq8B}). All solutions are obtained for initial conditions $r|_{\tau=0}=1$, $r_s|_{\tau=0}=1.3$, $L|_{\tau=0}=0.15$, $n|_{\tau=0}=1$, $A=10^{-3}$, $p_0 = 5\cdot 10^{-3}$, $q=1$, $B=2.5\cdot 10^{-3}$.}
\label{fg5}
\end{figure}
Taking into account this connection, let us describe evolution of large pore by equation set (\ref{eq29}) in dimensionless variables.
Now, with the account of condition $r_s \gg L$, let us estimate the parameter $\alpha$ (see (\ref{eq3B}) ) as:
\begin{equation}\label{eq5B} \alpha\approx\ \frac{r_s^2-r^2}{2L}  \end{equation}
From here, using the definition (\ref{eq6}6), let us find bispherical coordinates of pore and granule boundaries $\eta_{1,2}$:
\begin{equation}\label{eq6B} \quad \eta_1  =\textrm{arsinh} \left(\frac{\alpha}{r}\right) \approx \ln\left(\frac{r_s^2-r^2}{rL}\right),\quad
\eta_2 = \textrm{arsinh}\left(\frac{\alpha}{r_s}\right) \approx \ln\left(\frac{r_s^2-r^2}{r_sL}\right) \end{equation}
Then we find the difference $\eta_1-\eta_2 =\ln \left(\frac{r_s}{r}\right) $, and, correspondingly, made estimate of series sums:	
\begin{equation}\label{eq7B}
\Phi_1= \frac{r^2 L}{(r_s-r)(r_s^2-r^2)}\cdot F(\mu^2, \mu; \mu^3, \nu),\; \Phi_2= \frac{r r_s L}{(r_s-r)(r_s^2-r^2)}\cdot F(\mu^2, \mu; \mu^3, \widetilde \nu),
\end{equation}
where $F$ is hyper-geometric function, and the following designations are introduced $$ \mu=\frac{r_s}{r}, \quad \nu= \left(\frac{rL}{r_s^2-r^2}\right)^2, \quad  \widetilde \nu= \left(\frac{r_sL}{r_s^2-r^2}\right)^2 . $$
The numerical estimate of hyper-geometric functions (\ref{eq7B}) for initial parameters $r|_{\tau=0}=1$, $r_s|_{\tau=0}=1.3$, $L|_{\tau=0}=0.15$ gives the value of the order of unity, thus, in the further consideration we take
$$ \Phi_1 \approx \frac{r^2 L}{(r_s-r)(r_s^2-r^2)}, \quad  \Phi_2 \approx \frac{r r_s L}{(r_s-r)(r_s^2-r^2)} .$$
Let us now turn to evolution regime after gas pressure in the pore $P_g$ reaches values of outer pressure $P_0$.
Then we substitute the expressions for $\Phi_1$ and $\Phi_2$  into equation set  (\ref{eq29}) and, with account of the condition of the smallness of pore displacement $L(\tau) \approx L_0$ and the performance of the conditions $A \ll r, A\ll r_s$,  obtain simplified evolution equation for large pore:
\begin{equation}\label{eq8B}
\begin{cases}
L(\tau)\approx L|_{\tau=0}=L_0        \\
r_s=\left(V+r^3 \right)^{1/3} \approx V^{\frac{1}{3}}\left(1+\frac{r^3}{3V}\right)  \\
\frac{d r}{d \tau}=-\frac{A\cdot e^{-p_0}}{qr^2}\cdot \frac{r_s+r}{r_s-r}  \\
\frac{d n}{d \tau}=\left(  \frac{p_0}{B}\cdot  r -\frac{n}{r^2} \right)\frac{r_s }{r_s-r}   \\
\end{cases}
\end{equation}
As it can be seen from the last equation of the set (\ref{eq8B}), during characteristic time $ \tau_0 \sim \frac{r^2 (r_s -r)}{r_s} \sim r^2$, quasistationary gas pressure is established in the pore $n/r^3 \approx p_0 /B$, and equation for  pore radius change $r(\tau)$  splits from the rest of equations. Its solution can be found as:
\begin{equation}\label{eq9B}
\int \frac{(V^{1/3}\cdot(1+r^3/3V)-r)r^2}{V^{1/3}\cdot(1+r^3/3V)+r} dr= \frac{r^3}{3} -6 V^{2/3}\int \frac{r^3 dr}{r^3 +3 V^{2/3}r + 3V}= $$
 $$- \frac{A}{q}e^{-p_0}\cdot (\tau-\tau_0)+\textrm{const}
\end{equation}
Integral in (\ref{eq9B}) is calculated exactly (see. Appendix ), then analytical solution can be found easily:
\begin{equation}\label{eq10B}
\tau-\tau_0=-\frac{q}{A} \cdot \exp \left(p_0\right)\cdot\left(F(r(\tau))-F(r(\tau_0))\right),
\end{equation}
where the designation is introduced
$$F(r(\tau))=\frac{r(\tau)^3}{3}-6 V^{2/3}f(r(\tau), r_1),    $$
The change of gas atoms number in the pore $n(\tau)$ is determined according to $n(\tau)=\frac{p_0}{B}\cdot r^3(\tau)$, using solution of (\ref{eq10B}). The relation (\ref{eq10B}) allows also to estimate the life time of a large pore:
\begin{equation}\label{eq11B}
\Delta\tau \approx \frac{q}{A}e^{p_0} \left(\frac{r(\tau_0)^3}{3}-6 V^{2/3} f(r(\tau_0),r_1)+ 6 V^{2/3} f(0,r_1)\right)
\end{equation}
Let us use for the estimate  $r(\tau_0)\approx 1$ and parameter values   $A=10^{-3}$, $p_0 = 5\cdot 10^{-3}$, $q=1$, $r|_{\tau=0}=1$, $r_s|_{\tau=0}=1.3$. Then we obtain pore life time $\Delta\tau \approx 83$, that agrees very well with the observed $\Delta\tau \approx 80$ in Fig. \ref{fg5}. In Fig. \ref{fg5}, analytical solution is indicated by dashed line. From here, the good agreement can be seen of  analytical solutions with those for the simplified  equation set (\ref{eq8B}). The last are shown in Fig.\ref{fg5} by solid line. The Both of these solutions also agree well with numerical solution of complete equation set (\ref{eq29}), that in Fig. \ref{fg5} is indicated by dashed-and-dotted line.

\section{Gas pore in the center of spherical granule}

Let us consider now degenerate case when gas pore is situated exactly in the center of spherical granule $(l=0)$. Evidently, the inequality $R<R_s$ is also met. In this case, it is convenient to use the spherical coordinate system. The boundary conditions for vacancy and gas concentrations remain the same and are determined by formulae (\ref{eq1})-(\ref{eq3}) correspondingly. The stationary diffusion equations determining vacancy and gas concentrations take on the simple form:
\begin{equation}\label{eq1a}	
\frac{d}{d r}\left(r^2\frac{d}{d r}\right) c = 0,\quad \frac{d}{d r}\left(r^2\frac{d}{d r}\right) c_g = 0
\end{equation}
Equations (\ref{eq1a}) are complemented with boundary conditions
\[  \quad c(r)|_{r=R}=c_R^v, \quad  c(r)|_{r=R_s}=c_{R_s}^v,\quad \quad c_g(r)|_{r=R}=c_R^g, \quad  c_g(r)|_{r=R_s}=c_{R_s}^g. \]
The relations for the concentrations are easily found from equations (\ref{eq1a}) 
\begin{equation}\label{eq2a}	c(r)=-\frac{C_1}{r}+C_2,  \quad c_g(r)=-\frac{C_3}{r}+C_4    \end{equation}
where $C_{1,2,3,4}$ are free constants that are determined by the boundary conditions.
The vacancy and gas fluxes ${\vec j}_v$ and  ${\vec j}_g$ are determined by the first Fick law \cite{16s}, then vacancy and gas fluxes on the unit surface area of pore and granule are equal, correspondingly, to:
\begin{equation}\label{eq3a}
\vec n \cdot \vec j_v|_{r=R}=-\frac{D}{\omega}\frac{\partial c}{\partial r}|_{r=R},\quad \vec n \cdot \vec j_v|_{r=R_s}=-\frac{D}{\omega}\frac{\partial c}{\partial r}|_{r=R_s}
\end{equation}
\begin{equation}\label{eq4a}
\vec n \cdot \vec j_g|_{r=R}=\frac{D_g}{\omega}\frac{\partial c_g}{\partial r}|_{r=R}
\end{equation}
Substituting these expressions into the equations fore pore and granule volume change (\ref{eq16})-(\ref{eq17}), as well as into the equations foe the change of the number of gas atoms inside the pore(\ref{eq18}), one obtains
\begin{figure}
	\centering
	\includegraphics[width=5.5 cm, height=5.5 cm]{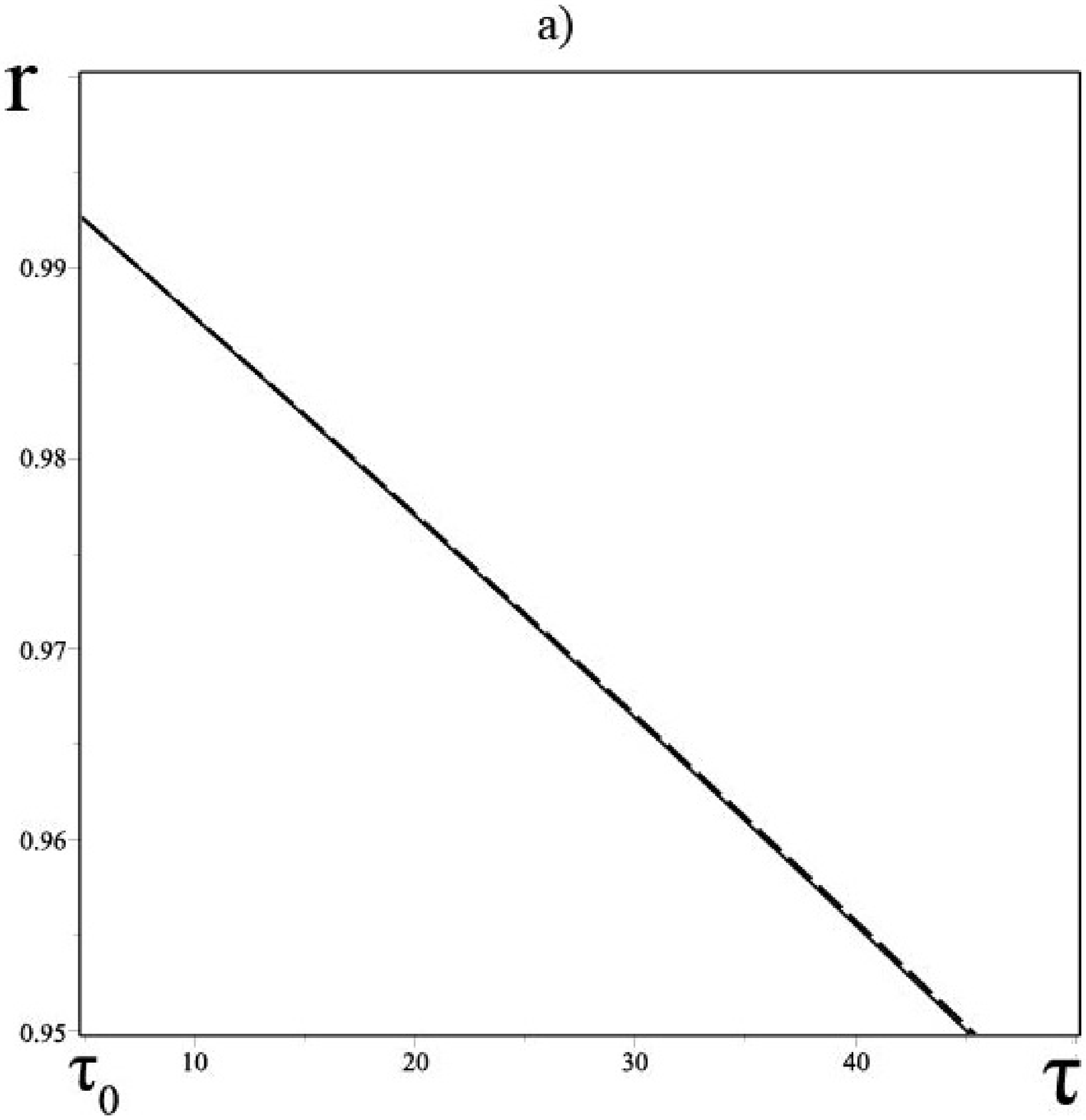}
	\includegraphics[width=5.5 cm, height=5.5 cm]{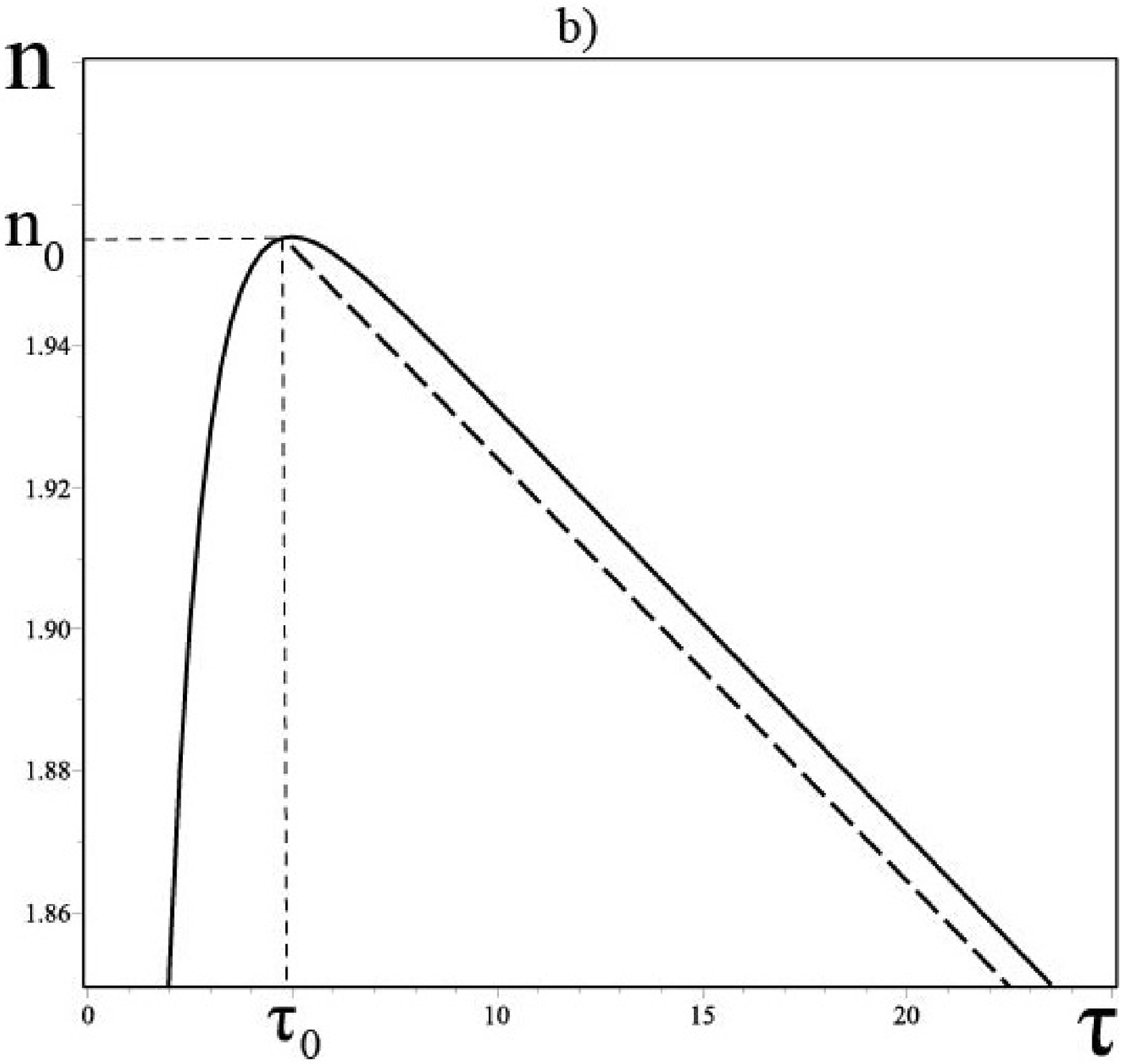}
	\includegraphics[width=5.5 cm, height=5.5 cm]{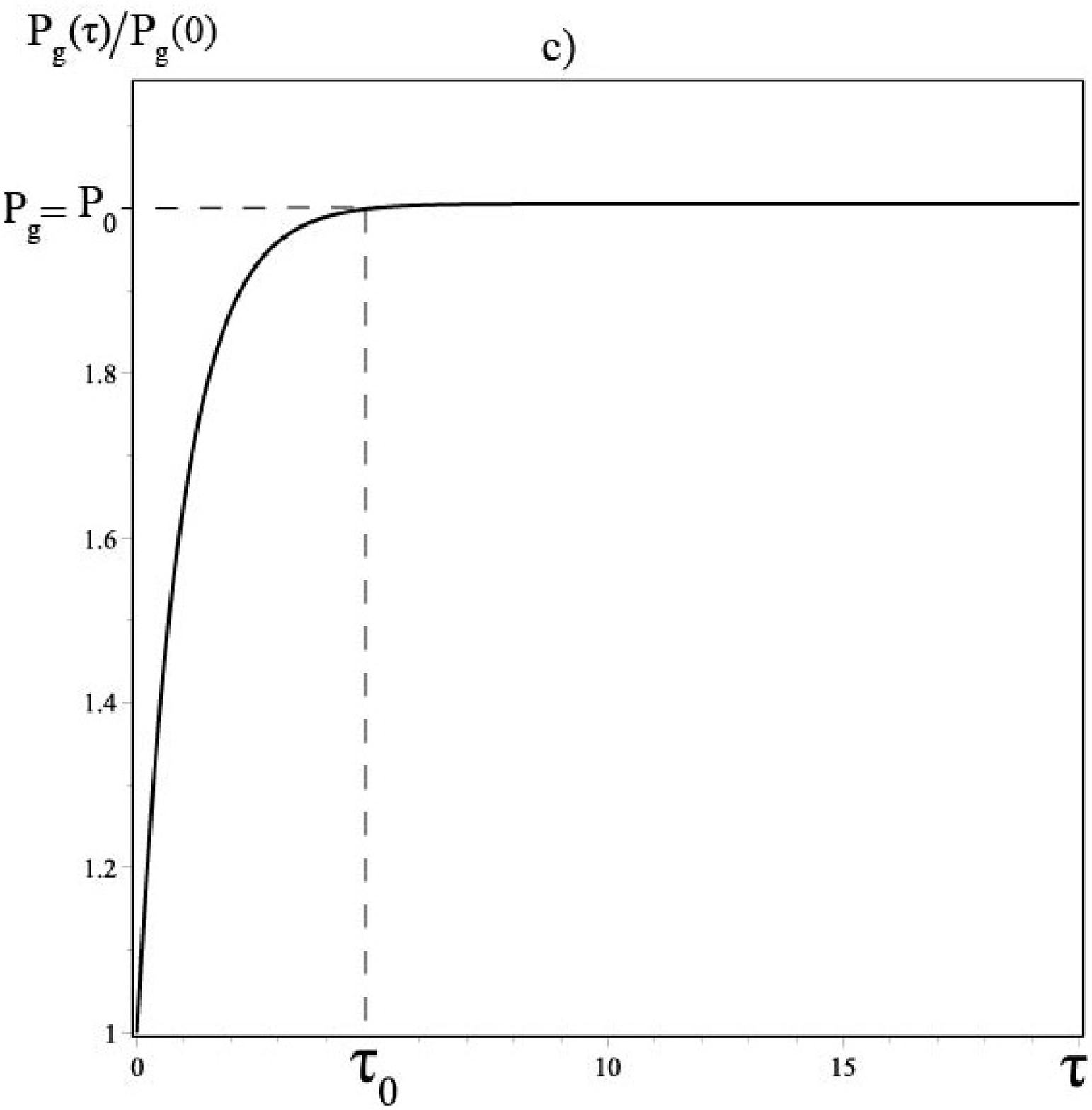}\\
	\caption{a) -- the plot of dependence of pore radius $r$ on time $\tau$; b) -- the plot of dependence of gas atom number in the pore $n$ on time $\tau$; c) -- the plot of dependence of gas pressure in the pore $P_g(\tau)/P_g(0)$ on time $\tau$. Solid line indicates the numerical solution of equation system  (\ref{eq9a}), while dashed line indicates analytical solutions of equations (\ref{eq9a}). All solutions are obtained for initial conditions $r|_{\tau=0}=1$, $r_s|_{\tau=0}=100$, $n|_{\tau=0}=1$, $A=10^{-3}$, $p_0 = 5\cdot 10^{-3}$, $q=1$, $B=2.5\cdot 10^{-3}$.}
	\label{fg6}
\end{figure}
the following equations:
\begin{equation}\label{eq5a}
\left\{
\begin{aligned}
\dot{R}=\frac{D}{R}\cdot\frac{(c_{R_s}^v-c_R^v)R_s}{R_s-R} \\
\dot{R_s}=\frac{D}{R_s}\cdot\frac{(c_{R_s}^v-c_R^v)R}{R_s-R} \\
\dot{N_g}=\frac{4\pi D_g}{\omega}\cdot\frac{(c_{R_s}^g-c_R^g)RR_s}{R_s-R} \\
\end{aligned}
\right.
\end{equation}
One can verify easily that, from the evolution equations (\ref{eq5a}) for gas pora in the granule center, follows the conservation low for granule material
\[R_s(t)^2\dot{R}_s(t)-R(t)^2\dot{R}(t)=0.\]
Then granule radius is connected with pore volume by simple relation
\begin{equation}\label{eq6a}
R_s(t) =\sqrt[3]{V+R(t)^3},
\end{equation}
where $V=R_s(0)^3 -R(0)^3$ is initial volume of granule material. Taking into account the relation (\ref{eq6a}),  gas pore evolution in the granule center is described by two differential equations:
\begin{equation}\label{eq8a}
\left\{
\begin{aligned}
\dot{R}=\frac{Dc_VR_s}{R(R_s-R)}\cdot\left[\exp\left(-\frac{2\gamma\omega}{kTR_s}-\frac{P_0\omega}{kT}\right)-\exp\left(\frac{2\gamma\omega}{kTR}-\frac{P_g\omega}{kT}\right)\right], \\
\dot{N_g}=\frac{4\pi D_gR_sR }{\omega(R_s-R)}\cdot \kappa_H \left[P_0-P_g\right], \\
P_g=\frac{N_gkT}{\frac{4\pi}{3}R^3}, \quad P_0=\textrm{const}, \quad R_s(t) =\sqrt[3]{V+R(t)^3} .\\
\end{aligned}
\right.
\end{equation}
with initial conditions: $R(0)=R_0, R_s(0)=R_{s0}, N_g(0)=N_0$.
Now let us turn to numerical solution of (\ref{eq8a}) in dimensionless form:
\begin{equation}\label{eq9a}
\begin{cases}
\frac{d r}{d \tau}=\frac{r_s}{qr(r_s-r)}\cdot\left[e^{-\frac{A}{r_s}-p_0}-e^{\frac{A}{r}-\frac{B}{r^3}\cdot n} \right] \\
\frac{d n}{d \tau}=\frac{r_s\cdot r}{r_s-r}\cdot\frac{p_0-\frac{B}{r^3}\cdot n}{B} \\
r_s=\sqrt[3]{V+r^3},\; p_0=\textrm{const}, \\
r|_{\tau =0}=1,  \; r_s|_{\tau =0}=100,\\
n|_{\tau =0}=1.
\end{cases}
\end{equation}
Here, as before, dimensionless variables are determined by expressions(\ref{eq28}).

When gas pressure in the pore $P_g$ reaches the value of external pressure $P_0$, equations (\ref{eq9a}) can be solved analytically. Indeed, implementing in the first equation of the system (\ref{eq9a}) the relation $n(\tau)/r^3(\tau)=p_0/B$, one obtains evolution equation for pore radius:
\begin{equation}\label{eq10a}
\frac{d r}{d \tau}=-\frac{\exp\left(\frac{A}{r}-p_0\right)}{qr}+\frac{\exp\left(-\frac{A}{r_{s0}}-p_0\right)}{qr}
\end{equation}
It is easy now to obtain general solution of (\ref{eq10a}) in integral form
\begin{equation}\label{eq11a}
\int \limits_{r_{0}}^r \frac{r d r}{1-\exp(A/r+A/r_{s0})}=\frac{\exp\left(-\frac{A}{r_{s0}}-p_0\right)}{q}\cdot (\tau-\tau_0)
\end{equation}
We conduct integration in (\ref{eq11a}) under the following conditions: small gas pores ($r \ll r_{s0}$) for which $A \ll r$ and $A \ll r_s$:
\begin{equation}\label{eq12a}
r(\tau)=\left(r^3(\tau_0)-\frac{3A}{q}\cdot \exp(-p_0) \cdot (\tau-\tau_0)\right)^{\frac{1}{3}}
\end{equation}
Moreover, pore  life time on this stage $\Delta \tau$ can be estimated as
\[\Delta \tau = \frac{r^3(\tau_0)q}{3 A}e^{p_0}\]
Using relation (\ref{eq9aa}), one finds the expression for the change of the number of atoms in the pore since the time moment $\tau_0$:
\begin{equation}\label{eq13a}
n(\tau)=\frac{p_0}{B}\cdot \left(r^3(\tau_0)-\frac{3A}{q}\cdot \exp\left(-p_0\right) \cdot (\tau-\tau_0)\right)
\end{equation}
In Fig. \ref{fg6}, The dependences are shown of radius $r(\tau)$, atoms number $n(\tau)$ and pressure $P_g(\tau)/P_g(0)$ in gas-filled pore on time  $\tau$. In Fig. \ref{fg6} solid line indicates numerical solutions of  equation set (\ref{eq9a}), while dashed line indicates analytical ones of (\ref{eq12a}) and (\ref{eq13a}). The good agreement of numerical and analytical results can be seen. All calculations were conducted for the following parameter values: $r|_{t=0}=1$, $r_s|_{t=0}=100$, $n|_{t=0}=1$, $A=10^{-3}$, $p_0 = 5\cdot 10^{-3}$, $q=1$, $B=2.5\cdot 10^{-3}$.

\section{Conclusions}

The behavior of gas-filled pore has been studied in a particle of limited size in gas atmosphere under the influence of diffusion fluxes of single atom gas and vacancies. The main attention is paid to the case, when initial gas pressure in the pore is smaller then outer gas pressure  $P_0>P_g(0)$. It is shown, that in general case,  gas pressure in the pore is increased up to the outer one. The further  pore evolution occurs at stationary  gas pressure in the pore and diminishing of pore radius. The number of gas atoms in the pore, at  initial stage, grows until gas pressure in the pore reaches the value of the outer one. After that, the number of gas atoms in the pore decreases while ore shifts towards granule center. Asymptotic modes of small and large pores were considered, that allowed to obtain simplified sets of nonlinear equations and find their analytical solutions. Life times of gas-filled pore were found  analytically for all asymptotic modes. These life times agree well with those obtained via numerical solution of complete equation set. In the case when initial gas pressure in the pore exceeds outer one $P_0<P_g(0)$, the stage is observed of diminishing  gas pressure in the pore down to the value of outer one while  pore radius grows and pore shifts towards granule boundary. After reaching by gas pressure in the pore the value of outer one, pore evolution occurs in the described above way.

One should take into account, that the case of molecular gas is essentially different form the considered case of single atom gas. The cause of this is that diffusion of atoms that constitute the molecule can occur with different diffusion coefficient. After reaching pore boundary these atoms form volatile molecular compound. This is just this molecular gas that fills in the pore. This case will be considered in the further work.

\section*{Appendix}

\subsection*{1.Auxiliary expressions.}
\begin{equation}\label{EQ1}
\int_{-1}^1\frac{P_k(t)dt}{\sqrt{\cosh\eta-t}} = \frac{\sqrt{2}\cdot
e^{-(k+1/2)\eta}}{k+1/2}\,.
\end{equation}
 Via differentiating the relation (\ref{EQ1}) over
parameter $\eta$, one consequently finds
\begin{equation}\label{EQ2}
\int_{-1}^1\frac{P_k(t)dt}{(\cosh\eta-t)^{3/2}} =
\frac{2\sqrt{2}\cdot e^{-(k+1/2)\eta}}{\sinh\eta}\,,
\end{equation}
\begin{equation}\label{EQ3}
\int_{-1}^1\frac{P_k(t)dt}{(\cosh\eta-t)^{5/2}} =
\frac{4\sqrt{2}\cdot
e^{-(k+1/2)\eta}(\cosh\eta+(k+1/2)\sinh\eta)}{3\cdot\sinh^3\eta}\,.
\end{equation}
\subsection*{2. Obtaining pore radius change.}

\begin{equation}\label{EQ4} \dot{R}=-\frac{\omega}{4\pi
R^2}\oint\vec{n}\cdot\vec{j}_v|_{\eta=\eta_1}dS\,,
\end{equation}
 where
\begin{equation}\label{EQ5}
\vec{n}\cdot\vec{j}_v|_{\eta=\eta_1}=\frac {D}{\omega} \cdot
\frac{\cosh\eta_1-\cos\xi}{a}\frac{\partial
c}{\partial\eta}|_{\eta=\eta_1}\,,
\end{equation}
 \begin{equation}\label{EQ6}
 dS = \frac{a^2\cdot \sin\xi
d\xi d\varphi}{(\cosh\eta_1-\cos\xi)^2}\,,
\end{equation}
 After implementing Fubini's theorem, with account of the independence of $\xi$ and $\varphi$ the expression for $\dot{R}$ takes on a form
 
 \begin{equation}\label{EQ7}
  \dot{R}=-\frac{a\cdot D}{2\cdot
R^2}\int_0^{\pi}\frac{\partial
c}{\partial\eta}|_{\eta=\eta_1}\frac{\sin\xi d\xi
}{\cosh\eta_1-\cos\xi}\,,
\end{equation}
 Substituting
$$\frac{\partial
c}{\partial\eta}|_{\eta=\eta_1} = \sqrt{2}\left(
\frac{c_{R}^v\cdot\sinh\eta_1}{2\sqrt{{\cosh\eta_1-\cos\xi}}}\cdot\sum_{k=0}^{\infty}
P_k(\cos\xi)\exp(-\eta_1(k+1/2))+
\sqrt{{\cosh\eta_1-\cos\xi}}\times\right.$$
$$\left.\times\sum_{k=0}^\infty\frac{(k+1/2)\cdot
P_k(\cos\xi)}{\sinh(k+1/2)(\eta_1-\eta_2)}\left[c_{R}^v\cdot\cosh(k+1/2)(\eta_1-\eta_2)e^{-\eta_1(k+1/2)}
-c_{R_s}^v\cdot e^{-\eta_2(k+1/2)}\right] \right)$$
into the expression for the speed of pore radius change, and exchanging integration and summation signs on the strength of convergence of corresponding sums and integrals, after substituting $\cos\xi=t$, one obtains

$$ \dot{R}=-\frac{a\cdot D\sqrt{2}}{2\cdot
R^2}\left[\frac{c_{R}^v\cdot\sinh\eta_1}{2}\sum_{k=0}^\infty
e^{-\eta_1(k+1/2)}\int_{-1}^1\frac{P_k(t)dt}{(\cosh\eta_1-t)^{3/2}}+\right.$$
$$\left. +\sum_{k=0}^\infty\frac{(k+1/2)}{\sinh(k+1/2)(\eta_1-\eta_2)}\left[c_{R}^v\cdot\cosh(k+1/2)
(\eta_1-\eta_2)e^{-\eta_1(k+1/2)} -\right.\right. $$
$$\left.\left.- c_{R_s}^v\cdot
e^{-\eta_2(k+1/2)}\right]\cdot
\int_{-1}^1\frac{P_k(t)dt}{\sqrt{\cosh\eta_1-t}}\right] \,.$$
Using values of integrals (\ref{EQ1}) and (\ref{EQ2}), we can reformulate this expression
$$ \dot{R}=-\frac{a\cdot D\sqrt{2}}{2\cdot
R^2}\left[\frac{c_{R}^v\cdot\sinh\eta_1}{2}\sum_{k=0}^\infty
e^{-\eta_1(k+1/2)}\cdot \frac{2\sqrt{2}\cdot
e^{-\eta_1(k+1/2)}}{\sinh\eta_1} +\right.$$
$$\left. +\sum_{k=0}^\infty\frac{(k+1/2)}{\sinh(k+1/2)(\eta_1-\eta_2)}\left[c_{R}^v\cdot\cosh(k+1/2)
(\eta_1-\eta_2)e^{-\eta_1(k+1/2)} -\right.\right. $$
$$\left.\left.- c_{R_s}^v\cdot
e^{-\eta_2(k+1/2)}\right]\cdot \frac{\sqrt{2}\cdot
e^{-(k+1/2)\eta_1}}{k+1/2} \right]= -\frac{a\cdot D}{R^2}\left[
c_{R}^v\cdot\sum_{k=0}^\infty e^{-\eta_1(2k+1)} +\right.$$
$$\left. +\sum_{k=0}^\infty \frac{c_{R}^v\cdot\cosh(k+1/2)
(\eta_1-\eta_2)e^{-\eta_1(k+1/2)} + c_{R_s}^v\cdot e^{-\eta_2(k+1/2)}}
{\sinh(k+1/2)(\eta_1-\eta_2)}\cdot e^{-\eta_1(k+1/2)}\right] =
-\frac{a\cdot D}{R^2}\times$$
$$\times\left[\frac{c_{R}^v}{2\cdot\sinh\eta_1} +\sum_{k=0}^\infty
\frac{c_{R}^v\cdot\cosh(k+1/2) (\eta_1-\eta_2)\cdot e^{-\eta_1(k+1/2)} -
c_{R_s}^v\cdot e^{-\eta_2(k+1/2)}} {\sinh(k+1/2)(\eta_1-\eta_2)}\cdot
e^{-\eta_1(k+1/2)}\right]\,.$$
Substituting  $a=R\cdot \sinh\eta_1$ and transforming summing terms,\\ 
we ultimately obtain: 
\begin{equation}\label{EQ8} \dot{R}=-\frac{D}{R}\left[\frac{c_{R}^v}{2}
+\sinh\eta_1\cdot\sum_{k=0}^\infty
\frac{c_{R}^v\cdot(e^{-(2k+1)\eta_1}+e^{-(2k+1)\eta_2}) -2\cdot
c_{R_s}^v\cdot e^{-(2k+1)\eta_2}}
{e^{(2k+1)(\eta_1-\eta_2)}-1}\right]\,.
\end{equation}

\subsection*{3. Calculation of the change of gas atoms number in the pore.}

\begin{equation}\label{EQ9} \dot{N_g}=\oint\vec{n}\cdot\vec{j}_g|_{\eta=\eta_1}dS\,,
\end{equation}
 where
\begin{equation}\label{EQ10}
\vec{n}\cdot\vec{j}_g|_{\eta=\eta_1}=-\frac {D_g}{\omega} \cdot
\frac{\cosh\eta_1-\cos\xi}{a}\frac{\partial
c_g}{\partial\eta}|_{\eta=\eta_1}\,,\quad dS = \frac{a^2\cdot \sin\xi d\xi d\varphi}{(\cosh\eta_1-\cos\xi)^2}\,
\end{equation}
From (\ref{eq12}) one finds the expression for
\begin{equation}\label{EQ11}
\frac{\partial c_g}{\partial\eta}|_{\eta=\eta_1} = \sqrt{2}\left(\frac{c_{R}^g\cdot\sinh\eta_1}{2\sqrt{{\cosh\eta_1-\cos\xi}}}\cdot\sum_{k=0}^{\infty}P_k(\cos\xi)\exp(-\eta_1(k+1/2))+\sqrt{{\cosh\eta_1-\cos\xi}}\times\right.$$
$$\left.\times\sum_{k=0}^\infty\frac{(k+1/2)\cdot P_k(\cos\xi)}{\sinh(k+1/2)(\eta_1-\eta_2)}\left[c_{R}^g\cdot\cosh(k+1/2)(\eta_1-\eta_2)e^{-\eta_1(k+1/2)}-c_{R_s}^g\cdot e^{-\eta_2(k+1/2)}\right] \right)
\end{equation}
Substituting (\ref{EQ11}) into (\ref{EQ10}), on the strength of independence of variables $\xi$ and $\varphi$, after implementing the substitution $\cos\xi=t$, one obtains:
$$ \dot{N_g}=-\frac{2\sqrt{2}\cdot\pi a \cdot D_g}{\omega}\left[\frac{c_{R}^g\cdot\sinh\eta_1}{2}\sum_{k=0}^\infty e^{-\eta_1(k+1/2)}\int_{-1}^1\frac{P_k(t)dt}{(\cosh\eta_1-t)^{3/2}}+\right.$$
$$\left. +\sum_{k=0}^\infty\frac{(k+1/2)}{\sinh(k+1/2)(\eta_1-\eta_2)}\left[c_{R}^g\cdot\cosh(k+1/2)(\eta_1-\eta_2)e^{-\eta_1(k+1/2)} -\right.\right. $$
$$\left.\left.- c_{R_s}^g\cdot e^{-\eta_2(k+1/2)}\right]\cdot \int_{-1}^1\frac{P_k(t)dt}{\sqrt{\cosh\eta_1-t}}\right] \,.$$
 Using values of integrals (\ref{EQ1}), (\ref{EQ2}) and $a=R\cdot \sinh\eta_1$, after easy transformations one obtains the final simple form of $\dot{N_g}$:
\begin{equation}\label{EQ12} \dot{N_g}=-\frac{4\pi D_g}{\omega}\cdot R\cdot\left[\frac{c_{R}^g}{2}
+\sinh\eta_1\cdot\sum_{k=0}^\infty
\frac{c_{R}^g\cdot(e^{-(2k+1)\eta_1}+e^{-(2k+1)\eta_2}) -2\cdot
c_{R_s}^g\cdot e^{-(2k+1)\eta_2}}
{e^{(2k+1)(\eta_1-\eta_2)}-1}\right]\,.
\end{equation}

\subsection*{4. Calculation of the speed of pore motion .}

$$ \vec{v}=\vec{e_z}\cdot\frac{3\cdot D\cdot
a}{2\cdot R^2}\int_0^\pi \frac{\partial
c}{\partial\eta}|_{\eta=\eta_1}\frac{\cosh\eta_1\cdot
\cos\xi-1}{(\cosh\eta_1-\cos\xi)^2}\cdot\sin\xi d\xi = $$
$$=\vec{e_z}\cdot\frac{3\cdot D\cdot
a}{2\cdot R^2}\int_0^\pi \frac{\partial
c}{\partial\eta}|_{\eta=\eta_1}\left(-\frac{\cosh\eta_1}{\cosh\eta_1-\cos\xi}+
\frac{\sinh^2\eta_1}{(\cosh\eta_1-\cos\xi)^2}\right) \cdot\sin\xi
d\xi =$$
$$=\vec{e_z}\cdot\frac{3\sqrt{2}\cdot D\cdot
a}{2\cdot R^2}
\int_0^\pi\left(-\frac{\cosh\eta_1}{\cosh\eta_1-\cos\xi}+
\frac{\sinh^2\eta_1}{(\cosh\eta_1-\cos\xi)^2}\right) \cdot\sin\xi
d\xi\times$$
$$\times
\left[ \frac{c_R^v\cdot\sinh\eta_1}{2\cdot\sqrt{\cosh\eta_1-\cos\xi}}\cdot\sum_{k=0}^\infty P_k(\cos\xi)
e^{-\eta_1(k+1/2)} + \sqrt{\cosh\eta_1-\cos\xi}\times\right.$$

$$\left.\times\left( \sum_{k=0}^\infty\frac{(k+1/2)
P_k(\cos\xi)(c_R^v\cdot e^{-\eta_1(k+1/2)}\cosh(k+1/2)(\eta_1-\eta_2) -c_{R_s}^v\cdot
e^{-\eta_2(k+1/2)})}{\sinh(k+1/2)(\eta_1-\eta_2)}
 \right)\right]\,.$$
The substitution $t=\cos\xi$ and exchange of summation and integration signs yield the expression
$$\vec{v}=\vec{e_z}\cdot\frac{3\sqrt{2}\cdot D\cdot
a}{2\cdot R^2}
\sum_{k=0}^\infty\left[\int_{-1}^1\frac{P_k(t)dt}{(\cosh\eta_1-t)^{5/2}}\cdot
\frac{c_R^v\cdot e^{-\eta_1(k+1/2)}\sinh^3\eta_1}{2}
\right.+$$
$$+\int_{-1}^1\frac{P_k(t)dt}{(\cosh\eta_1-t)^{3/2}}\times\left(\frac{-c_R^v\cdot e^{-\eta_1(k+1/2)}
\cosh\eta_1\sinh\eta_1}{2}\right.+\sinh^2\eta_1\times$$
$$\left.
\times\frac{(k+1/2)
(c_R^v\cdot e^{-\eta_1(k+1/2)}\cosh(k+1/2)(\eta_1-\eta_2) -c_{R_s}^v\cdot
e^{-\eta_2(k+1/2)})}{\sinh(k+1/2)(\eta_1-\eta_2)}
 \right)+\int_{-1}^1\frac{P_k(t)dt}{\sqrt{\cosh\eta_1-t}}\times$$
$$ \left.\times\left(
-\cosh\eta_1\cdot\frac{(k+1/2)
(c_R^v\cdot e^{-\eta_1(k+1/2)}\cosh(k+1/2)(\eta_1-\eta_2) -c_{R_s}^v\cdot
e^{-\eta_2(k+1/2)})}{\sinh(k+1/2)(\eta_1-\eta_2)}
 \right)\right]\,.$$
Now let us substitute values of corresponding integrals \\ into the obtained expression and reform the result
$$\vec{v}=\vec{e_z}\cdot\frac{3\sqrt{2}\cdot D\cdot
a}{2\cdot R^2} \sum_{k=0}^\infty\left[\frac{4\sqrt{2}\cdot
e^{-(k+1/2)\eta_1}(\cosh\eta_1+(k+1/2)\sinh\eta_1)}{3\cdot\sinh^3\eta_1}\times
\right. $$
 $$\times\frac{c_R^v\cdot e^{-\eta_1(k+1/2)}\sinh^3\eta_1}{2}+\frac{2\sqrt{2}\cdot
e^{-(k+1/2)\eta_1}}{\sinh\eta_1}\times
\left(\frac{-c_R^v\cdot e^{-\eta_1(k+1/2)}\cosh\eta_1}{2}
+\sinh^2\eta_1\times\right.$$
$$\left. \times\frac{(k+1/2)
(c_R^v\cdot e^{-\eta_1(k+1/2)}\cosh(k+1/2)(\eta_1-\eta_2) -c_{R_s}^v\cdot
e^{-\eta_2(k+1/2)})}{\sinh(k+1/2)(\eta_1-\eta_2)}
 \right)+\frac{\sqrt{2}\cdot e^{-(k+1/2)\eta_1}}{k+1/2}\times$$
$$\left. \times\left(
-\cosh\eta_1\cdot\frac{(k+1/2)
(c_R^v \cdot e^{-\eta_1(k+1/2)}\cosh(k+1/2)(\eta_1-\eta_2) -c_{R_s}^v\cdot
e^{-\eta_2(k+1/2)})}{\sinh(k+1/2)(\eta_1-\eta_2)}
 \right)\right]=$$
\begin{equation}\label{EQ13}=\vec{e_z}\cdot\frac{3 D
a}{ R^2}
\sum_{k=0}^\infty\frac{((2k+1)\sinh\eta_1-\cosh\eta_1)\left(c_R^v\cdot(e^{-(2k+1)\eta_1}+e^{-(2k+1)\eta_2})-2c_{R_s}^v\cdot
e^{-(2k+1)\eta_2}\right)}{e^{(2k+1)(\eta_1-\eta_2)}-1}\,.
\end{equation}

\subsection*{5. Calculation of the integral in (\ref{eq9B}).}

Cubic polynomial in the denominator of the expression (\ref{eq9B}) can be easily presented in the necessary form.  Equality of this polynomial to zero has one real root
\[r_1 = V^{1/3} \left( \frac{(-12+4 \sqrt{13})^{1/3}}{2}  - \frac{2}{(-12+4 \sqrt{13})^{1/3}} \right)\]
and, correspondingly, this polynomial can be brought to the form
\[r^3+3 r V^{2/3}+3 V=(r-r_1)(r^2 +r_1 r + r_1^2 +3V^{2/3})\]
Integral (\ref{eq9B})  is calculated exactly if transformed to the form
\[\int  \frac{x^3 dx}{(x-a)(x^2+a x+a^2 +3 V^{2/3})}=x+{\frac {{a}^{3}\ln  \left( x-a \right) }{3\,{a}^{2}+3\,{V}^{2/3}}}-1
/2\,{\frac {{a}^{3}\ln  \left( {x}^{2}+ax+{a}^{2}+3\,{V}^{2/3}
 \right) }{3\,{a}^{2}+3\,{V}^{2/3}}}-
\]
\[-18\,\arctan \left( {\frac {2\,x+a
}{ \sqrt{3\,{a}^{2}+12\,{V}^{2/3}}}} \right) {V}^{2/3}{a}^{2} \left( 3
\,{a}^{2}+3\,{V}^{2/3} \right) ^{-1} \left(  \sqrt{3\,{a}^{2}+12\,{V}^
{2/3}} \right) ^{-1}-\]
\[-18\,\arctan \left( {\frac {2\,x+a}{ \sqrt{3\,{a}^
{2}+12\,{V}^{2/3}}}} \right) {V}^{4/3} \left( 3\,{a}^{2}+3\,{V}^{2/3}
 \right) ^{-1} \left(  \sqrt{3\,{a}^{2}+12\,{V}^{2/3}} \right) ^{-1}-\]
\[-3
\,\arctan \left( {\frac {2\,x+a}{ \sqrt{3\,{a}^{2}+12\,{V}^{2/3}}}}
 \right) {a}^{4} \left( 3\,{a}^{2}+3\,{V}^{2/3} \right) ^{-1} \left(
 \sqrt{3\,{a}^{2}+12\,{V}^{2/3}} \right) ^{-1} \equiv f(x,a)\]


\begin{thebibliography}{17}

\bibitem{1s} Slezov V. V. The coalescence of the system of dislocation loops and pore under irradiation, Solid State Physics, 1967, vol. 9, No. 12, p.3448.
\bibitem{2s} Slezov V.V., Shikin V.B. Coalescence of pores at presence of bulk vacancy sources, Euro nuclears., 1965, vol.2,  No. 3, pp.127-131.
\bibitem{3s} Saralidze Z.K., Slezov V.V. Coalescence of dislocation loops at nonstationary regime, Solid State Physics, 1965, vol. 7, No. 3, p.1605.
\bibitem{4s}  V.V. Slezov, V.V. Sagalovich,  Diffusive decomposition of solid solutions, Sov. Phys. Usp.30 (1987), pp. 23-45.
\bibitem{5s}  Slezov V.V. The gasfilled pore onset in solid solutions, Solid State Physics, 1995, vol.37, No.10, pp. 2879-2891.
\bibitem{6s}  Slezov V.V., Osmaev O.A.,  Shapovalov R.V. Poremotion in material with a source of gas atoms. Questions of atomic science and technique. 2005.  No. 3. Volume: Physics of radiation damage and radiation material science (86), pp. 38-42.
\bibitem{6ss} V. V. Slezov, A. S. Abyzov, Zh. V. Slezova, The Nucleation of Gas-Filled Bubbles in Low-Viscosity Liquids, Colloid Journal, 2004, vol. 66, Issue 5, pp. 575–583. 
\bibitem{7s}  P.G. Cheremskoy, V.V. Slyozov, V.I. Betehin, Pores in Solid Matter, Energoatomizdat, Moscow, 1990 (in Russian). 
\bibitem{7ss} Alan J. Markworth, On the coarsening of gas-filled pores in solids, Metallurgical Transactions, 1973, vol. 4, Issue 11, pp. 2651–2656.
\bibitem{8s}  Problems of solid state theory/ NAS of Ukraine: KPTI. editors: Baryahktar V.G., Pletminskiy S.V. et al. -- Kiev: Naukova dumka, 1991, 200 p.
\bibitem{9s}  Saralidze Z.K., Slezov V.V., On the theory of coalescence of gas pores. Solid State Physics, 1965, vol.7, No.6, pp. 1605-1611
\bibitem{10s} Slezov V.V. Theory of gas pore growth at diffusion decomposition of multicomponent systems. Metallofizika, 1981, vol. 3, No.1, pp. 21-29.
\bibitem{10ss} V. V. Slezov, Gas Bubbles in Viscous Liquids and Melts, Journal of Colloid and Interface Science, 2002, vol. 255, Issue 2, pp. 274-292.
\bibitem{11s} Ragulya A.V., Skhorohod V.V., Consolidation of nanostructural materials, Kiev, Naukova dumka, 2007, 374p.
\bibitem{11ss} V.I. Dubinko, A.V. Tur, A.A. Turkin and V.V. Yanovsky, Diffusion interaction of new-phase precipitates at random distances, Phys. Met. Metallogr. Vol.68. (1989), pp.17-25.
\bibitem{12s} Y. Yin, R. M. Rioux, C. K. Erdonmez, S. Hughes, G. A., A. P.  Formation of hollow nanocrystals through the nanoscale Kirkendall effect. Science, 304 (2004), pp.711-714.
\bibitem{13s} Zaporozhets T.V., Gusak A.M., Podolyan O.N. Evolution of pore in nanoshells –- competition of direct and reverse Kirkindale effect, effects of Frenkel and Gibbs-Tomphson (phenomenological description and computer simulation).  Usp. Fiz. Met.2012, vol. 13, pp.1-70.
\bibitem{14s} V. V. Yanovsky , M. I. Kopp, M. A. Ratner. Evolution of vacancy pores in bounded particles. arXiv:1809.06565v1[cond-mat.mes-hall] (2018)
\bibitem{15s} V.V. Yanovsky, M.I. Kopp, M. A. Ratner.  Evolution of gas-filled pore in bounded particles. arXiv:1810.103319v1[cond-mat.mes-hall]  (2018)
\bibitem{16s} Ja.E.  Geguzin, M.A. Krivoglaz, Motion of Macroscopic Inclusions in Solid Matter, Metallurgy, Moscow, 1971 (in Russian).
\bibitem{17s} G. Arfken, Mathematical Methods in Physics, Atomizdat, Moscow, 1970 (in Russian). 


\end{thebibliography}
\end{document}